\documentclass[a4paper,11pt]{article}
\usepackage[scale={0.75,0.8}]{geometry}
\pdfoutput=1 

\usepackage{jcappub} 

\pdfoutput=1
\usepackage[utf8]{inputenc}
\usepackage{microtype}
\usepackage{times}
\usepackage{graphicx}
\usepackage{subfig}
\usepackage{hyperref, cleveref}
\usepackage{amsmath,bm}
\usepackage{color}
\usepackage{epstopdf}
\usepackage{float}
\usepackage{upgreek}
\usepackage{amssymb}
\usepackage[thinc]{esdiff}
\usepackage{cleveref}
\usepackage{soul}
\crefname{equation}{Eq}{Eqs} 
\crefrangelabelformat{equation}{(#3#1#4--#5#2#6)}
\crefname{figure}{Fig}{Figs} 
\crefrangelabelformat{figure}{(#3#1#4--#5#2#6)}
\usepackage{pdflscape}
\usepackage[nice]{nicefrac}
\usepackage[dvipsnames]{xcolor}
\usepackage[figuresleft]{rotating}
\usepackage[numbers, sort&compress]{natbib}
\usepackage{makecell}
\usepackage{tikz}
\usetikzlibrary{positioning}
\usepackage{enumerate}
\usepackage{comment}
\def\Mp{M_{\rm Pl}}

\def\dx{\mathrm{d}}
\def\weff{w_{\mathrm{eff}}}

\def\neff{N_{\mathrm{eff}}}
\def\mf{m_{\phi}}

\def\phiin{\Phi_{\mathrm{in}}}

\definecolor{pred}{RGB}{190,0,100}
\definecolor{pblue}{RGB}{0,0,190}

\newcommand\abs[1]{\left|#1\right|}

\usepackage{eso-pic}
\newcommand{\reportnum}[2]{
  \AddToShipoutPictureBG*{%
    \AtPageUpperLeft{%
      \hspace{0.75\paperwidth}%
      \raisebox{#1\baselineskip}{%
        \makebox[0pt][l]{\textnormal{#2}}
  }}}%
}

\title{
Potential Surge Preheating: enhanced resonance from potential features
}

\author[a]{Pankaj~Saha}
\author[a,b,c,d]{~and~Yuko Urakawa}

\affiliation[a]{Institute of Particle and Nuclear Studies~(IPNS), High Energy Accelerator
Research Organization (KEK), Oho 1-1, Tsukuba 305-0801, Japan}
\affiliation[b]{The Graduate University for Advanced Studies (SOKENDAI), Tsukuba 305-0801, Japan}
\affiliation[c]{International Center for Quantum-field Measurement Systems for Studies of the Universe
and Particles (QUP), High Energy Accelerator Research Organization (KEK), Tsukuba,
Ibaraki 305-0801, Japan}
\affiliation[d]{Kobayashi-Maskawa Institute for the Origin of Particles and the Universe, Nagoya University, Nagoya 464-8602, Japan}

\emailAdd{pankaj@post.kek.jp}
\emailAdd{yukour@post.kek.jp}

\abstract{
We investigate the effects of local features in the inflationary potential on the preheating dynamics after inflation. We show that a small feature in the potential can enhance the resonance and bring the radiation-like state equation during preheating despite the inflationary potential being a quadratic one.
Such localized features may naturally arise due to various physical effects without altering the large-scale predictions of the original model for cosmic microwave background (CMB) observables.
We demonstrate that these features effectively introduce localized higher-power terms in the potential, significantly influencing the preheating dynamics~---~a phenomenon we term potential surge preheating.
We outline the resulting modifications in energy distribution among different components.
We further show that these small-scale features leave detectable imprints in the form of gravitational wave signals. These signals influence CMB measurements of the effective number of relativistic species, $N_{\mathrm{eff}}$, offering a way to reconstruct the shape of the inflaton potential at small scales.
Finally, we argue that these modifications to the scalar potential provide a framework to explore preheating dynamics and the fragmentation of scalar fields using simple scalar potentials.
}

\graphicspath{{./figs/}}

\begin{document}

\reportnum{-3}{KEK-TH-2678}
\reportnum{-4}{KEK-Cosmo-0370}

\maketitle
\flushbottom


\section{\label{sec:introd}Introduction}
The inflationary Universe is the leading candidate for explaining the `\textit{initial conditions}' for the standard cosmological big-bang evolution~\cite{Starobinsky:1980te,Guth:1980zm,Sato:1980yn,Linde:1981mu,Albrecht:1982wi,Linde:1983gd}.
Inflation provides the initial seed for large-scale structure formation from quantum vacuum fluctuations, providing us with many testable signatures on various cosmic microwave background (CMB) and gravitational waves (GWs) based experiments.
However, inflation is only a part of the story for unraveling the Big Bang puzzle. 
It was soon realized (almost immediately after the conception of the inflationary universe) that, after the end of inflation, we need to transfer the energy locked in the inflaton to repopulate the Universe and commence the radiation-dominated (RD) universe. 
This critical phase in the Early Universe, between the end of inflation to the RD stage, when the inflation is supposed to decay non-perturbatively due to resonance effects (at initial stages) and/or perturbatively (at the final stage) to other fields and radiation components, was dubbed as reheating~\cite{Abbott:1982hn,Dolgov:1982th,Albrecht:1982mp}. 
\par
The initial non-perturbative phases of reheating, when the inflaton exists as a coherent condensate rather than a large collection of statistically independent particles, are collectively known as preheating. 
This scenario differs significantly from the later perturbative single-body decays.
In the simplest preheating scenarios, we have a light `daughter' field $\chi$ coupled to inflaton $\phi$ via a simple four-legged interaction $g^2\phi^2\chi^2$. 
Creating all bosonic particles in the Early Universe, including scalar fields as a prototype, can happen analogously as in this simple toy example via parametric resonance. 
The occupation number of the produced fields grows exponentially. The transfer of energy happens within a very short time span (compared to the perturbative processes)~\cite{Traschen:1990sw,Kofman:1994rk,Shtanov:1994ce,Kaiser:1995fb,Kofman:1997yn,Greene:1997fu,Kaiser:1997mp,Kaiser:1997hg}.
In addition to the couplings to the daughter field for decay, preheating also depends crucially on the shape of the inflaton potential.
Depending on the shape of the inflaton potential, explosive particle production can happen solely via self-resonance in the inflaton sector without any additional couplings~\cite{Lozanov:2016hid,Lozanov:2017hjm,DeCross:2015uza,DeCross:2016cbs,DeCross:2016fdz}. On the other hand, changing the interaction sector lead to a plethora of preheating scenarios; for instance, preheating involving trilinear, higher-order interactions and/or multiple daughter fields~\cite{Dufaux:2006ee,Croon:2015naa,Antusch:2015vna,Enqvist:2016mqj,Antusch:2021aiw,Antusch:2022mqv,Cosme:2022htl}.
Other variations such as geometric preheating where the fields are non-minimally coupled to gravity~\cite{Bassett:1997az,Tsujikawa:1999jh,Fu:2019qqe,Figueroa:2021iwm}, tachyonic preheating after hybrid inflation~\cite{Felder:2000hj,Felder:2001kt,GarciaBellido:2002aj,Copeland:2002ku} and multi-field (p)reheating scenarios can be found in~\cite{Bezrukov:2008ut,GarciaBellido:2008ycs,Braden:2010wd,Giblin:2010sp,DeCross:2015uza,DeCross:2016fdz,DeCross:2016cbs,Krajewski:2018moi,Iarygina:2018kee,Figueroa:2022iho,Krajewski:2022ezo}. The production of gauge fields during preheating~\cite{Rajantie:2000nj,Copeland:2001qw,Smit:2002yg,GarciaBellido:2003wva,Tranberg:2003gi,Skullerud:2003ki,vanderMeulen:2005sp,DiazGil:2007fch,DiazGil:2008raf,Dufaux:2010cf,Adshead:2015pva,Tranberg:2017lrx,Cuissa:2018oiw,Adshead:2019lbr,Cui:2021are} are also studied extensively. 
At this point, it is also worth noting that the gravitational backreactions and modes coupling are commonly neglected during preheating. 
This is a valid approximation due to the negligible influence of local gravity at the small-scale phenomena characteristic of this period~\cite{Frolov:2008hy,Huang:2011gf}.
Recent fully nonlinear numerical-relativistic field-theory simulations have extensively validated this simplification for both scalar and gauge field systems~\cite{Giblin:2019nuv,Kou:2019bbc,Joana:2022uwc,Aurrekoetxea:2023jwd,Adshead:2023mvt}.
\par
\begin{figure}[!ht]
    \centering
    \begin{minipage}{0.48\textwidth}
        \centering
        \includegraphics[width=0.68\textwidth]{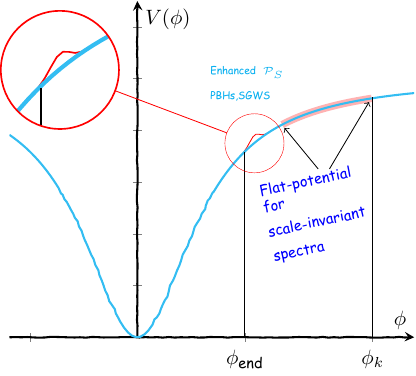} 
    \end{minipage}
    \begin{minipage}{0.48\textwidth}
        \centering
        \includegraphics[width=0.9\textwidth]{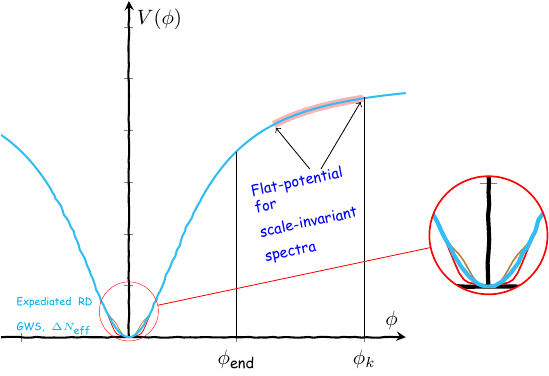} 
    \end{minipage}
    \caption{
    The CMB observables constrain only a portion of the field space of the inflationary potential where a flat potential is preferred. 
    A feature on the potential before the end of inflation but away from minima may enhance the small-scale power spectra with interesting phenomenological consequences, such as producing primordial black holes or sourcing scalar-induced secondary gravitational waves. 
    In this work, we expound on how some features around the minima of the potential affect the reheating phase. 
    Such features can also be discerned from the Gravitational waves generated from scalar field inhomogeneities. Furthermore, the features may emerge at different field intervals within a single inflationary potential, displaying distinct characteristics; therefore, combining all such observations at different scales could help us reconstruct the full inflaton potential.
    }
    \label{fig:schematic}
\end{figure}

In this work, we explore a different avenue: the effects of local small-scale
features in the inflationary potential on the preheating dynamics.
Such deformations in the potential on the intermediate scales (after the CMB scales but before the end of inflation) are widely studied in the literature in the context of generating enhanced power-spectrum that source the generation of primordial black holes (PBHs) and secondary gravitational waves (SGWs)~\cite{Starobinsky:1992ts,Adams:2001vc,Garcia-Bellido:2017mdw,Germani:2017bcs,Ballesteros:2017fsr,Hertzberg:2017dkh,Atal:2019cdz,Mishra:2019pzq,Kefala:2020xsx,Dalianis:2021iig,Inomata:2021tpx,Boutivas:2022qtl,Karam:2022nym,Gu:2022pbo,Inomata:2022yte,Pi:2022zxs}. (In this context, consider also the production of PBHs and SGWs from the collapse of small-scale perturbations arising from preheating instabilities, as explored in~\cite{Bassett:2000ha,Suyama:2004mz,Jedamzik:2010hq,Jedamzik:2010dq,Martin:2019nuw,Martin:2020fgl,Papanikolaou:2020qtd})
Typically, the primordial features are considered in the literature during inflation; consequently, those features appear on large field values compared to the end of inflation. (For reviews on primordial features, see~\cite{Chluba:2015bqa,Slosar:2019gvt}; for observational searches, see~\cite{Chen:2016vvw,Ballardini:2016hpi,Palma:2017wxu,LHuillier:2017lgm,Ballardini:2017qwq,Debono:2020emh,Braglia:2022ftm,Euclid:2023shr}).
Here, we show that such a local feature on a very small scale can have interesting consequences for preheating, too (see the schematic figure~\ref{fig:schematic}). 
We describe how the change in the resonance structure due to local deformations in the form of a dip or a bump near the minima of the scalar potential affects the preheating without altering the inflationary predictions for the model. 
Specifically, these small-scale features in the inflationary potential can alleviate certain challenges associated with scalar preheating and enable a more rapid transition to a radiation-dominated universe.
Furthermore, the nonlinear stage of preheating is known to produce GWs through the fragmentation of the coherent scalar field.
We demonstrate that these deformations in the scalar potential can also be directly detected through GW spectra and their influence on the effective number of relativistic degrees of freedom, $\Delta \neff$.
\par 
We have structured the remainder of the paper as follows: We will first outline the details of features in the base quadratic scalar potential in Sec~\ref{sec:model}.
In Sec~\ref{sec:preh_all}, we will describe the linear and nonlinear phases of the parametric resonance and non-perturbative dynamics induced by a coherent oscillating scalar field.
The results of the GWs production due to the fragmentation of the coherent scalar are collected in Sec~\ref{sec:sec_gws}. 
We conclude in Sec.~\ref{sec:disc_conc}.
We will consider $\hbar = c = 1$ unless otherwise stated. 
We have used $\Mp( = 1/\sqrt{8\pi G}=2.43\times10^{18}\mathrm{GeV})$ to denote the reduced Planck mass. 
We will take the usual Friedmann-Le\^{i}matre-Roberson-Walker (FLRW) metric as our background metric $\dx s^2= \dx t^2 - a^2(t)(\dx x^2+\dx y^2+\dx z^2) $, where $a(t)$ is the scale factor and $t$ being the cosmic time.

\section{\label{sec:model}Scalar potential with Gaussian features}
\begin{figure}[!ht]
    \centering
    \includegraphics[width=0.7\textwidth]{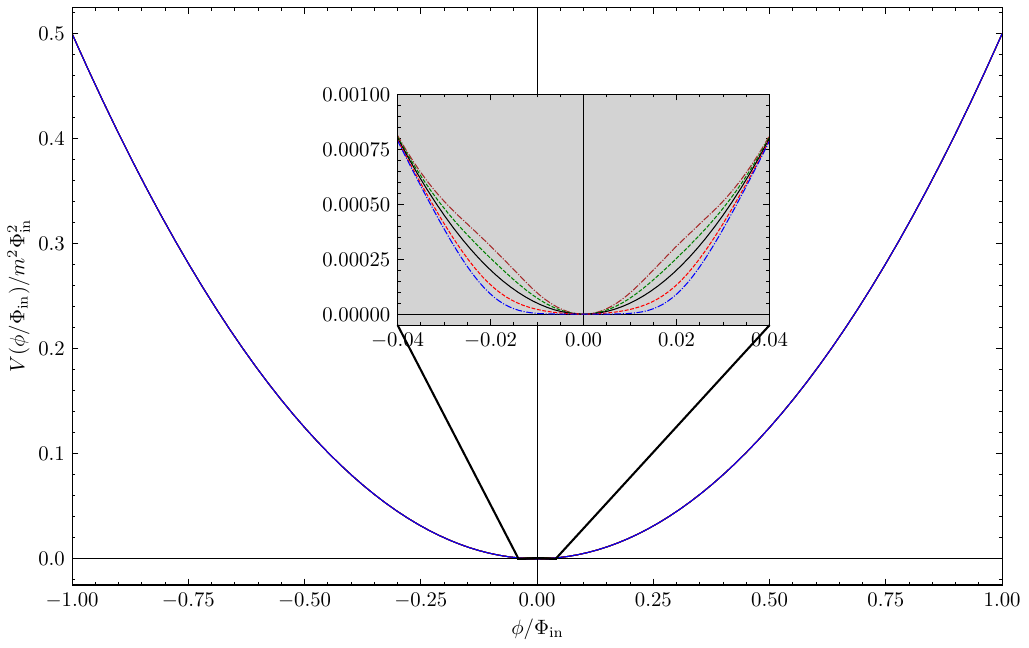}
    \caption{
    We plot the shape of the rescaled scalar potential in the dimensionless field variable during the preheating epoch for the following values of $h$: Black: $h=0$~(the vanilla $m^2\phi^2$), brown (dot-dashed) $h=0.815$, green (dashed) $h=0.4$, red (dashed) $h=-0.5$, and blue (dot-dashed) $h=-0.815$. The features are symmetrically placed at $\phi_S = 10^{-2}\Mp$. We also took the variances $\sigma = |\phi_S|$. The dimensionless field is defined in the units of the initial amplitude of the oscillation.
    The potential shape is modified locally, so the inflationary predictions for the models are not altered.}
    \label{fig:figVphi}
\end{figure}
In the inflationary universe scenario, the Universe was in a state of negative pressure when the scale factor grew exponentially in time.
A universe dominated by a scalar field displaced from its minima can provide such a state of negative pressure.
Various probes for inflation are directed toward probing the properties of the scalar field, including the shape of the scalar potential. 
CMB observations are sensitive to a limited region of the scalar potential's field space. 
In these regions, models with a nearly flat potential are favored by the data.
Away from this range, the potential shape is not constrained, and importantly, a deviation in those regions from the regular power-law behaviors has interesting observational signatures, such as in the form of PBHs or SGWs. 
To model such deviations in the scalar potential, let us start with the general form of the potential with a deformation function $\delta_i(\phi,\phi_i)$:
\begin{equation}
V(\phi) = V_0(\phi)\left(1 + \sum_i\delta_i(\phi)\right),
\label{eq:pot0}
\end{equation}
where $V_0(\phi)$ is the original potential responsible for generating the spectra compatible with observations at CMB scales. 
The local deformation can be in the form of a step~\cite{Adams:2001vc,Adshead:2011jq}, a bump/dip~\cite{Atal:2019cdz,Mishra:2019pzq}, or, in the form of an inflection point~\cite{Cicoli:2018asa} will change the shape of the function away from the scales probed by CMB~\footnote{See~\cite{Caravano:2024tlp} for a recent study on the effects of Small-scale features on the inflationary dynamics using lattice simulation.}. 
\par
Before delving further into the morphology of the potential, let us mention shortly why such deformations might arise from a more fundamental perspective~\cite{Bartolo:2013exa,Pi:2022ysn}. 
One such case could be the renormalization group (RG) effects, analogous to the case discussed in~\cite{Espinosa:2017sgp,Espinosa:2018eve}. 
Inflationary models in string theory can also give rise to a local deformation in the potential of a slow-roll inflationary models~\cite{Cicoli:2008gp,Cicoli:2018asa}.
However, in this work, we will not explore the origins of these deformations but rather focus on the phenomenological aspects of the deformations in the preheating phase.
To compare with the existing literature for preheating, we have used the ubiquitous $V_0(\phi) = \frac{1}{2}m^2\phi^2$ potential as our base potential to highlight the effects of the features on the preheating phase.
Preheating in many observationally preferred inflationary models, such as Starobinsky inflation and $\alpha$-attractor models, can be approximated by a quadratic potential near the minima.
Additionally, as we are mainly concerned about the relative effect of small deformation on the potential shape compared to the original one and how such features can alleviate the outstanding issue of preheating with massive inflaton, this simple model illustrates all essential features.
We will describe the effect of such features on other observationally favored models that reduce to $m^2\phi^2$ potential around the minima in the appendix.
Additionally, depending upon the coupling, the inflaton can decay into various daughter fields (scalars, fermions, or gauge fields).
The efficiency of decay and the subsequent process of thermalization depend both on the scalar potential and the nature of the interaction involved.
As emphasized in the introduction, we will consider the scale-free four-legs interaction $\mathcal{L}_{\mathrm{int}}\subseteq\frac{1}{2}g^2\phi^2\chi^2$ as our interaction term. 
Moreover, this type of interaction represents the leading term for scalar-gauge interactions\cite{Figueroa:2015rqa} and is safe if a reflection symmetry protects the inflaton.
Consequently, we will study the effect of preheating, considering a small deviation from the vanilla $m^2\phi^2$ model with a four-legged interaction to be our base model given as:
\begin{align}
   V_{\mathrm{tot}}(\phi,\chi) &= \frac{1}{2}m^2\phi^2\left(1 + \delta(\phi)\right) + \frac{1}{2}g^2\phi^2\chi^2,
   \label{eq:full_potential}
\end{align}
with the following Gaussian deformation:
\begin{align}
\label{eq:gdef1}
\delta(\phi) & = h\left(\exp\left(-\frac{1}{2}\frac{(\phi - \phi_{S})^2}{\sigma^2}\right) + \exp\left(-\frac{1}{2}\frac{(\phi + \phi_{S})^2}{\sigma^2}\right)\right),\\
\label{eq:gdef2}
&= 2h\exp\left(-\frac{\phi_{S}^2}{2\sigma^2}\right)\cosh\left(\frac{\phi_S~\phi}{\sigma^2}\right)\exp\left(-\frac{\phi^2}{2\sigma^2}\right),
\end{align}
where for $h>0$, it corresponds to bumps, and $h<0$, it corresponds to dips.
The Gaussian bumps/dips are at $\phi=\pm\phi_S$, and $\sigma$ determines the spread of the feature. 
Throughout this work, we choose the parameters $h$ and $\sigma$ so that the global minimum remains at $\phi = 0$. 
Fig~\ref{fig:figVphi} shows the potential shapes for the relevant range of $h$.
It is evident from Eq.~(\ref{eq:gdef2}) that the feature is a Gaussian-like term with a heavier tail due to the hyperbolic cosine term with the height being modulated by the combination $2\exp(-\phi_S^2/2\sigma^2)$.
The mean of the $\delta(\phi)$ is zero while its variance is found to be $\sigma_c^2 = \sqrt{2}(\sigma^2 + \phi_S^2)$. 
Thus, the full potential is still $Z_2$ symmetric while the appearance of the bumps/dips at $\phi=\pm\phi_S$ only modifies the dynamics around a few $\sigma_c$ without altering the dynamics during inflation.  
On the other hand, near the minima at $\phi=0$, the potential is again described by $\phi^2$ term with the mass term scaled by a factor $\left(1 + 2h\exp\left(-\frac{\phi_S^2}{2\sigma^2}\right)\right)$. 
For simplicity, we will set $\sigma = |\phi_S|$ for the rest of the work. 
With this basic morphology of the potential with features at hand, we will describe their effects on the preheating dynamics in the following sections.

\section{\label{sec:preh_all}Preheating for the deformed potentials}
The essential idea of particle production during preheating is the following~\cite{Kofman:1994rk}~---~the homogeneous inflaton $\phi$ starts oscillating coherently when its effective mass becomes comparable to the rate of Hubble expansion~---~This coherently oscillating field provides a classical time-dependent background on which any field coupled to it can be produced due to parametric resonance. 
A linear analysis is applicable during the initial stage of preheating and can capture some of the essential features of the resonance structure.
The amplification of the fluctuations and the draining of energy from homogeneous components soon make the linear analysis redundant.
The system enters a strong backreaction regime when scattering between different modes is no longer negligible.
We must solve the full nonlinear system in the lattice, which will be performed in Sec~\ref{sec:pre_nl}.
\subsection{\label{sec:pre_lin}Linear analysis for potential deformation}
Let us begin by writing the full action for the system of scalar fields $\phi$ and $\chi$ given by the potential in~(\ref{eq:pot0})
\begin{align}
S = \int \sqrt{\abs{g}}\dx^4 x \bigg[\frac{\Mp^2}{2}R - \frac{1}{2}g^{\mu\nu}\nabla_{\mu}\phi\nabla_{\nu}\phi - \frac{1}{2}g^{\mu\nu}\nabla_{\mu}\chi\nabla_{\nu}\chi - V_{\mathrm{tot}}(\phi,\chi) \bigg],
\label{eq:action}
\end{align}
where $V_{\mathrm{tot}}(\phi,\chi) = V(\phi) + \frac{1}{2}m_{\chi}^2\chi^2 + \frac{1}{2}g^2\phi^2\chi^2$. 
For simplicity, we will ignore the bare-mass for $\chi$.
The equations  of motion, in the FLRW metric, are:
\begin{align}
	\label{eq:phi}
	&\ddot{\phi} + 3H\dot{\phi} - \frac{\nabla^2\phi}{a^2} + \frac{\partial}{\partial\phi}V_{\mathrm{tot}}(\phi,\chi) =0,\\
	\label{eq:chi}
		&\ddot{\chi} + 3H\dot{\chi} - \frac{\nabla^2\chi}{a^2} + \frac{\partial}{\partial\chi}V_{\mathrm{tot}}(\phi,\chi) =0,
\end{align}
with the Hubble expansion $H=\dot{a}/a$ is given by the First of the Friedmann equations as
\begin{align}
\nonumber
      H^2 &= \frac{1}{3 \Mp^2}\rho_{\mathrm{tot}},\\
      &= \frac{1}{3 \Mp^2}\left[\frac{1}{2}\dot{\phi}^2 + \frac{1}{2}\dot{\chi}^2 +  \frac{1}{2a^2}(\nabla \phi)^2 + \frac{1}{2a^2}(\nabla \chi)^2 + V_{\mathrm{tot}}(\phi,\chi)\right].
      \label{eq:H}
\end{align}
We generally consider the inflaton to be initially homogeneous, while the field $\chi$ is initially not excited and represents a \emph{quantum} field being excited in the \emph{classical} time-dependent background.
For simple $m^2\phi^2$ potential, the background field has the following solution $\phi_0(t) \approx \Phi(t)\sin(mt)$ with the amplitude function $\Phi(t) \sim \Mp/mt \propto 1/a^{3/2}$. The field starts rolling/oscillating when $H(t_{\ast})\approx m$, which sets the initial amplitude of oscillation to be $\Phi(t_{\ast}) = \phiin$.
For the potential will small-scale features~---~due to the emergence of higher-power potential terms~---~we expect the oscillation to get modified from simple sinusoidal behavior.
For an oscillating scalar in a $V(\phi)\propto\phi^n$ potential, the amplitude varies with the scale factor as $\Phi\propto a^{-6/(n+2)}$ while the energy density scales as $\rho_{\phi}\propto a^{-2n/(n+2)}$. 
Consequently, the amplitude of inflaton, $\Phi(t)$, redshifts more slowly for a larger $n$, with which the interaction between $\phi$ and $\chi$ remains important for a longer duration.
The inflaton, in general, also has fluctuations $\phi(\vec{x}, t) = \phi_0(t) + \delta\phi(\vec{x},t)$, whose fluctuations can also grow exponentially until the backreaction becomes important.

\subsubsection{\label{sec:pre_pr}Parametric resonance}
Parametric resonance occurs when certain modes of a dynamic system, coupled with an oscillating background, experience exponential growth. This phenomenon arises when specific parameters, such as the coupling strength, fall within particular resonance bands.
Now, ignoring the backreaction of the produced particles, the mode equation for $\delta\phi_k(t)$ and $\chi_k(t)$, at the linear order, takes the following form 
\begin{align}
\label{eq:delphik}
\delta\ddot{\phi}_k + 3 H \delta \dot{\phi}_k + \left(\frac {k^2}{a^2} + \frac{\partial^2V}{\partial\phi_0^2} \right) \delta\phi_k &= 0,\\
\label{eq:chik}
\ddot{\chi}_k + 3 H \dot{\chi}_k + \left(\frac {k^2}{a^2} + g^2\phi_0^2 \right) \chi_k &= 0.
\end{align}
The associated number density of the particles with momentum $k$ is given by \cite{Kofman:1997yn}
\begin{equation}
n_{ k} = \frac{\omega_{k}}{2} \left(\frac{|\dot{f_{ k}}|^2}{\omega_{ k}^2} + |f_{ k}|^2\right) - \frac{1}{2}\,,
\end{equation}
where $\omega_k^2 = \frac{k^2}{a^2} + \frac{\partial^2V}{\partial f^2}$ and $f=(\phi,\chi)$.
\par
The analytic study of parametric resonance has been developed using simple power-law chaotic potentials in the pioneering works in~\cite{Kofman:1994rk,Greene:1997fu} (See also, \cite{Boyanovsky:1994me,Boyanovsky:1995ema,Boyanovsky:1996sq,Boyanovsky:1996rw,Kaiser:1994vs,Kaiser:1995fb,Kaiser:1997hg,Kaiser:1997mp}). Meanwhile, for any arbitrarily complicated potentials, we may need to solve the equations above numerically. 
A qualitative understanding of the preheating instability can be achieved by casting the mode equation into the well-known Mathieu/Hill type differential equation~\cite{McLachlan:1947,Magnus:1966}. 
To proceed with this, we use the following rescaled variables
\begin{equation}
\tilde{\phi} = \frac{\phi}{\phiin},\quad\tilde{\chi} = \frac{\chi}{\phiin},\quad d\tilde{t} = m dt,
\label{eq:rescaled_vars}
\end{equation}
Now, for linear analysis, neglecting the universe expansion~\footnote{This is a good approximation well after the onset of the oscillation.}, Eqs.~(\ref{eq:delphik})-(\ref{eq:chik}) can be written as:
\begin{align}
\label{eq:delphik_hill}
\delta\tilde{\phi}_k'' + \left({\kappa^2} + \frac{\partial^2\tilde{V}}{\partial\tilde{\phi}_0^2} \right) \delta\tilde{\phi}_k &= 0,\\
\label{eq:chik_hill}
\tilde{\chi}_k'' + \left({\kappa^2} + \frac{g^2\tilde{\phi}_0^2}{m^2} \right) \tilde{\chi}_k &= 0,
\end{align}
where the prime denotes a derivative with respect to the dimensionless time variable $\tilde{t}$ and $\kappa = k/m$.
We have also used the dimensionless inflation potential as $\tilde{V}=V/(m^2\phiin^2)$. With these definitions, we can define the resonance parameter for $\chi_k$ as $q = g^2\Phi^2(t)/m^2$ with the initial resonance parameter given by $q_{\mathrm{in}}=g^2\phiin^2/m^2$. 
The nature of this oscillation depends on the form of the inflationary potential.
For the present case of potential with features, we can have self-resonance in the inflaton fluctuations due to the self-interaction induced by the bump or dip terms.
In case of resonance, the solutions of Eqs.~(\ref{eq:delphik_hill}) and (\ref{eq:chik_hill}) are marked by
an exponential growth $\delta\phi_k,\chi_{\kappa} \propto e^{\mu_k z}$ with $\mu_k$'s being the respective Floquet exponents that measure the strength of resonance. 
\par
\begin{figure}[!ht]
    \centering
    \begin{minipage}{0.48\textwidth}
        \centering
        \includegraphics[width=0.9\textwidth,height=5cm]{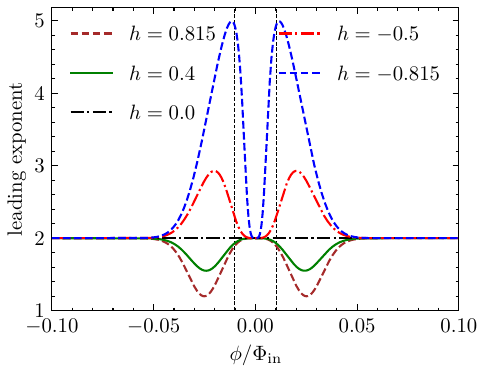} 
    \end{minipage}
    \begin{minipage}{0.48\textwidth}
        \centering
        \includegraphics[width=0.9\textwidth,height=5cm]{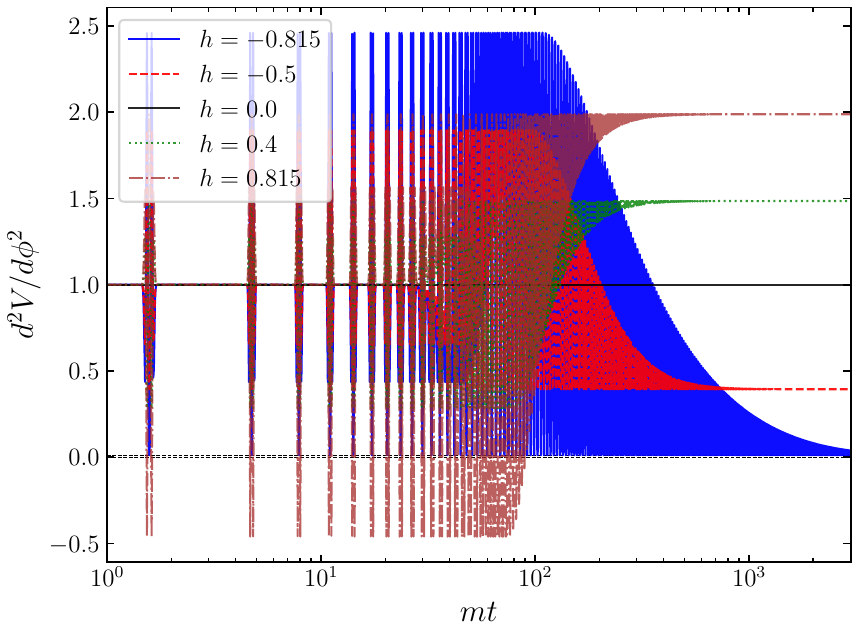} 
    \end{minipage}
    \caption{
    We show the 
    departure from the quadratic behavior of the full potential around the position of the features at $\phi_S=10^{-2}\Mp$. 
    The position of the features is shown using the vertical black dashed lines.
    In the right panel, we plot the evolution effective mass of background inflaton ($d^2V/d\phi^2$) with time.
    }
   \label{fig:BG01}
\end{figure}

To discern the strength and scale of the emergent higher-power terms, we plot the local slope of the $\log\phi$-$\log V$ to find the leading exponent of the potential as shown in Fig~\ref{fig:BG01}.
As we can see from Fig~\ref{fig:BG01}, the potential will be dominated by higher power potential terms $n>2$ due to the presence of the bumps. On the other hand, the potential will be shallower than quadratic around the position of the features when $h>0$.
The emergence of these local terms changes the nature of oscillation and resonance structure. 
For instance, during the coherent oscillation when the equation of state in a simple $V(\phi)\propto|\phi|^n$ type potential is given by $w=(n-2)/(n+2)$. Consequently, the EOS will quickly rise to a value greater than $1/3$ due to the higher-power potential terms.
The inflaton enters the potential region with the bumps/dips several oscillations after the onset of the oscillation. For $h< 0$, the potential is dominated by higher power terms with $n>2$ around $\phi=\phi_S$, resulting in a slower reduction of the oscillation amplitude $\Phi(t)$ compared to the quadratic case. In contrast, for $\phi>0$, the potential is dominated by terms smaller than quadratic $n<2$ that will reduce the amplitude faster.
This larger amplitude extends the duration of resonance for the $\chi$ field, driven by the $\phi$ oscillations, making it similar to the resonance in the standard $\phi^n$ with varying $n>2$.
The resonance parameter $q$ scales with expansion as $q(a) = q_{\mathrm{in}}a^{-12/(n+2)}$. 
Consequently, decreasing $h$ will result in an extended period of resonance for $\chi$ in potentials with bumps due to a slower transition from \emph{broad} to \emph{narrow} resonance bands and a slower reduction in background $\phi$ amplitude.
Next, let us discuss the self-resonance, which is absent for simple $m^2\phi^2$ potential at the linear level and requires further discussion.
For $h>0$, the resonance structure is similar to the preheating for $\phi^n$ with $n<2$, since the potential locally mimics to this case around $\phi=\pm\phi_S$.
\begin{figure}
    \centering
    \includegraphics[width=0.85\linewidth]{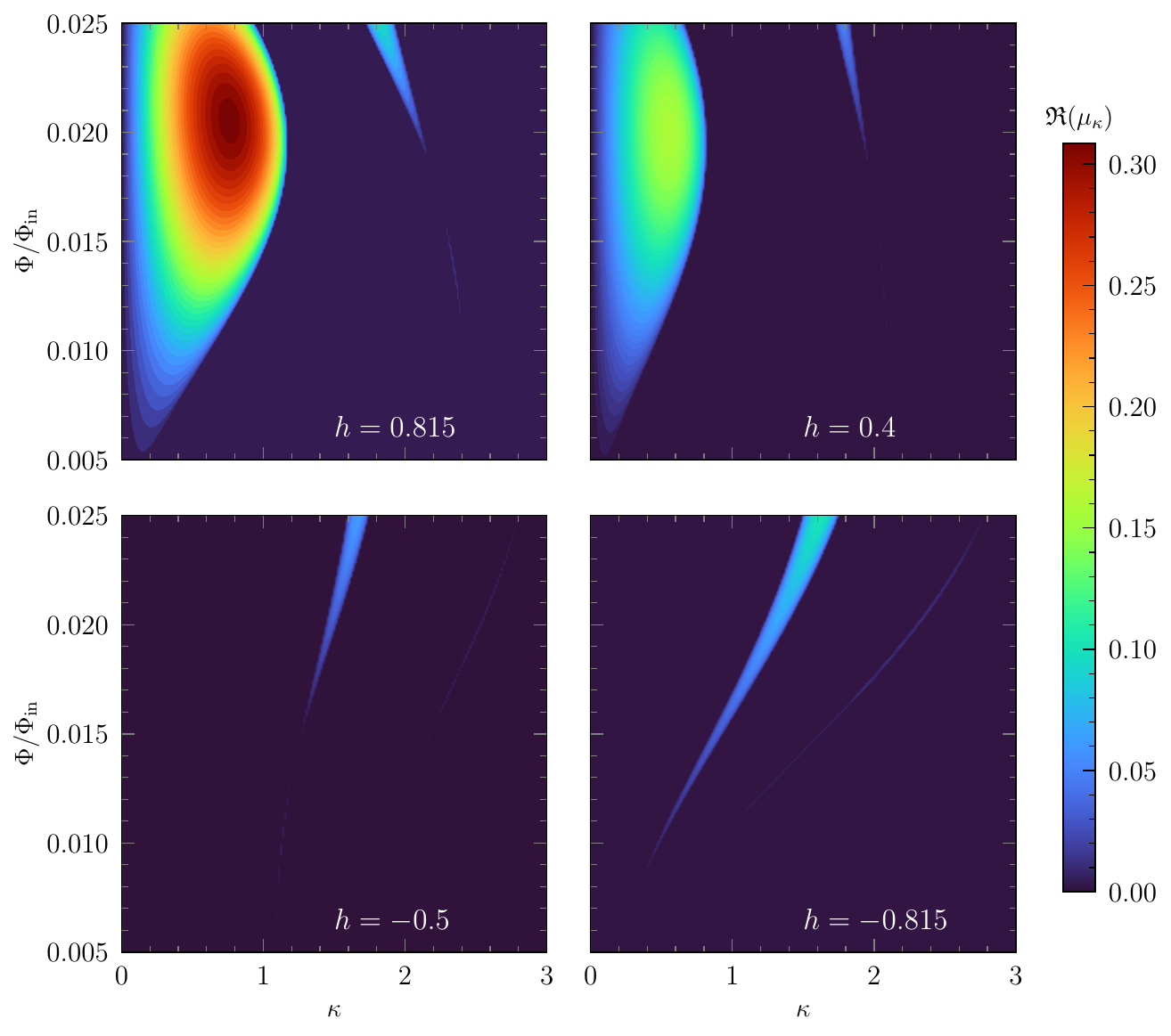}
    \caption{Instability diagram for self-resonance for different values of $h$. For $h>0$, when the potentials are steeper than quadratic, the self-resonance is very strong; the resonance structure resembles that of the plateau models. On the other hand, for $h<0$ due to the emergent higher power terms, the resonance structure is similar to $V(\phi)\propto \phi^n$ potentials with $n\geq 2$.}
    \label{fig:figFloquetSelfRes}
\end{figure}
For $h<0$, the structure mirrors $\phi^n$ with $n>2$ potentials. 
To understand the effect of the height of the Gaussian feature on the resonance structure, we plot the instability regions (real part Floquet exponent $\mathfrak{R}(\mu_k)>0$) for the $\delta\phi_k$ sector in Fig~\ref{fig:figFloquetSelfRes}.
The self-resonance increases as we move away from $h=0$ in both directions, which is due to the increasing non-linearity of the potential and can be seen from the growing strength of resonance encoded in $\mathfrak{R}(\mu_k)$.
\par
Within the broader context of preheating, the key distinction between negative and positive $h$ is that for $h>0$, the terms higher than quadratic contribute to the potential that redshift slowly than radiation, complicating the transition to a radiation-dominated (RD) Universe. The inflaton oscillates in a progressively narrower potential, resulting in a faster decrease in the oscillation amplitude, reducing the resonance in the $\chi$-sector. In these cases, preheating is largely driven by the self-resonance in inflaton fluctuations, leading to a system described by single-field dynamics. Consequently, a positive $h$ may favor the formation of exotic solitonic states but is not favorable for transitioning to an RD Universe.
In contrast, the potential widens for $h<0$, allowing the inflaton to oscillate with a larger amplitude for an extended period, resulting in slower but more sustained resonance in the $\chi$ field. 
Additionally, Fig~\ref{fig:BG01} illustrates another helpful fact that assists the preheating process further. 
For preheating with a massive component, as the energy of the daughter field redshifts faster than the massive $\phi$ component, the system tends to go back to a complete $\phi$-component domination eventually. 
We observed that introducing the features can lead to the surge of higher-power terms in the potential. These terms can eventually help us slow down the emergence of the $\phi$-dominated phase at a later stage.
However, preheating is typically dominated by the daughter field resonance when external couplings are present. 
Nonetheless, self-resonance in quadratic inflationary potentials with features will be an interesting phenomenon. We will consider the details of self-resonance in later sections. 

\subsection{\label{sec:pre_nl}Non-perturbative dynamics of the scalar field and lattice}
The exponential growth of field fluctuations due to parametric resonance rapidly approaches the scale of background energy densities. 
As a result, the linear analysis of the previous section becomes inadequate when backreaction from generated fields and mode interactions intensify.
We need to use lattice simulation to capture these effects with the expansion of the Universe calculated self-consistently. 
We use the publicly available lattice simulation code $\bm{\mathcal{C}\texttt{osmo}\mathcal{L}\texttt{attice}}$~\cite{Figueroa:2021yhd} (See also~\cite{Figueroa:2020rrl} for the details of art and science of the lattice simulation) for this purpose. 
Our \emph{program variables} for solving the equations in Eqs.~(\ref{eq:phi})-(\ref{eq:H}) on a lattice coincides with the choice made in Eqs.~(\ref{eq:rescaled_vars}) for linear analysis earlier with the spatial dimensions similarly rescaled as $d\tilde{x}^i = mdx^i$.
We have used at least $256^3$ lattice points for most of our simulations in a $3+1$ lattice. Some simulations are done with $512^3$ lattice points for convergence. 
Since the inflationary dynamics and the end of inflation are not altered with the introduction of the features, we start our simulation when the homogeneous field is $\phiin = 0.965\Mp$ for all our simulations.  
The mass parameter $m$ in such models is chosen to give the correct amplitude of the scalar power spectrum on CMB scales yielding $m = 5\times 10^{-6}\Mp$. 
To ensure that the inflationary potential remains unaffected by radiative corrections due to the interaction $g^2\phi^2\chi^2$ at CMB scales, we will take the $g^2~\sim 10^{-8}$ for all such cases. This choice will fix our resonance parameter as $q_{\mathrm{in}} = 10^4$ for the two-field preheating.
We will now illustrate the various features of the potential surge preheating in the subsequent sections.

\subsubsection{Background solutions and backreaction}
\begin{figure}[!ht]
    \centering
    \begin{minipage}{0.48\textwidth}
        \centering
        \includegraphics[width=0.9\textwidth]{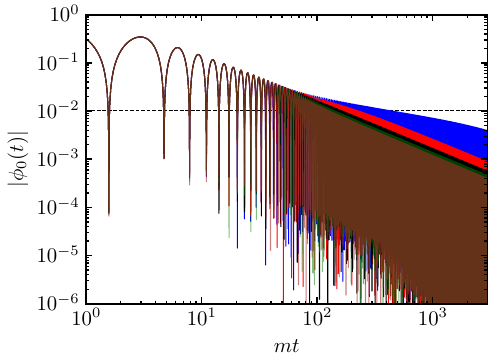} 
    \end{minipage}
    \begin{minipage}{0.48\textwidth}
        \centering
        \includegraphics[width=0.9\textwidth]{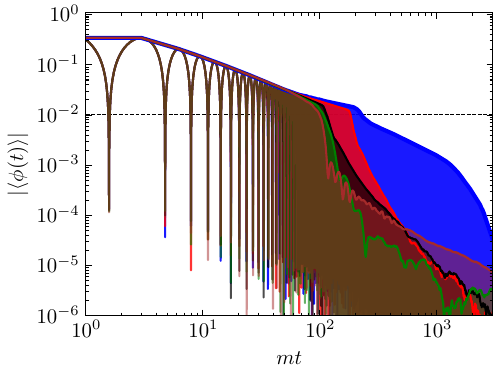} 
    \end{minipage}
    \caption{The variation of inflaton amplitude (in units of $\phiin$) with time for different values of $h$ (from blue, green, black, red and cyan for $h=-0.815,\,-0.5,\,0.0,\,0.4,$ and $0.815$, respectively).
    The left panel shows amplitude change without the second field $\chi$ and without taking the backreaction, which shows the effect of the potential shape on the inflaton amplitude.
    The figure on the right is from the full lattice simulation with interaction.
    The dashed horizontal line shows the position of the features. 
    We have also plotted the amplitude envelope in the right panel to track the decrease in oscillation amplitude.
    }
    \label{fig:lna_vs_f0}
\end{figure}
The background oscillations govern the resonant behavior and energy distribution during the preheating for a given scalar field model.
Therefore, it will be natural to ask how the potential features affect the background oscillation of $\phi$.
As the Universe expands, the inflaton amplitude diminishes, rendering the resonance ineffective once it falls below a critical threshold.
For instance, in preheating with vanilla $m^2\phi^2$ type potential and four-legs interaction, the resonance becomes inefficient as the initial amplitude drops below $\Phi_{\ast} \lesssim m/g (=\phiin/\sqrt{q_{\mathrm{in}}})$~\cite{Kofman:1997yn}. 
For the chosen resonance parameter $q_{\mathrm{in}} =10^4$, this critical amplitude $\Phi_{\ast} = 0.01\phiin$ roughly coincides with the position of the bumps/dips in the potential. 
In addition to the expansion, the inflaton amplitude decreases further due to backreaction from the created particles. 
Thus, to ensure sufficient preheating, we need to choose a sufficiently large value of the resonance parameter such that it maintains a value larger than unity for the broad resonance to endure before the backreaction kicks in.
We have shown the evolution of inflaton amplitude with dimensionless time parameter in Fig \ref{fig:lna_vs_f0}. 
The left panel of the figure is the solution to Eq.~(\ref{eq:phi}) without the second field $\chi$ and neglecting the gradient term (the homogeneous inflaton equation). 
The right panel shows the lattice result of the full system, with the angular brackets representing the spatial average.
In the case of the amplitude with a full lattice solution (right panel), the backreaction from the produced particles fragments the coherently oscillating scalar, resulting in a faster reduction in the background field amplitude.
\par
The variations in the inflaton amplitude with the height of the features, without backreaction, are consistent with the typical amplitude scaling for $\phi^n$ models. As a result, during the domination of the higher-power potential terms, the field oscillates with larger amplitude, albeit with lesser energy density.
However, when backreaction kicks in, it reduces the amplitude significantly.
The backreaction causes the amplitude to decay within a similar time frame for most values of $h$. However, for sufficiently large bumps, such as $h=-0.815$, the oscillation amplitude remains sizable for a longer duration.
\begin{figure}[!ht]
     \centering
     \includegraphics[width=0.7\linewidth]{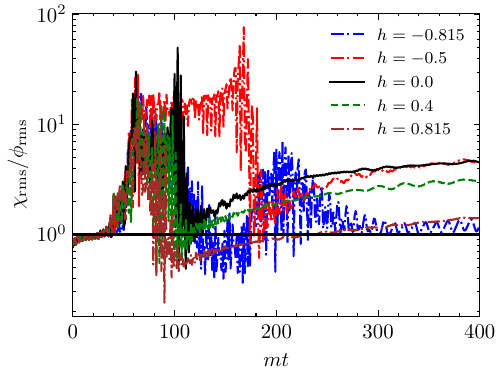}
     \caption{
     We plot the ratio of $\chi_{\mathrm{rms}}$ to $\phi_{\mathrm{rms}}$.
     The initial increase in the ratio is due to the parametric resonance in the $\chi$.
     When $h$ is away from zero, the strong self-resonance causes the ratio to achieve a value close to one.
     The generation of $\delta\phi$ from rescattering also decreases the ratio even when $h=0$. However, a strong self-resonance keeps this ratio close to one for a longer duration.
     }
     \label{fig:rms_ratio}
 \end{figure}
To understand the effect of backreaction further, we can check the evolution of the RMS values of the field: $\phi_{\mathrm{rms}} \equiv \sqrt{\langle\phi^2\rangle-\langle\phi\rangle^2}$ and $\chi_{\mathrm{rms}} \equiv \sqrt{\langle\chi^2\rangle-\langle\chi\rangle^2}$. 
In addition, the growth of field fluctuations seen in the respective RMS also indicates the particle production of the corresponding species~\cite{Amin:2014eta}.
In particular, we plot the ratio of $\chi_{\mathrm{rms}}/\phi_{\mathrm{rms}}$ as shown in Fig~\ref{fig:rms_ratio}.
The initial phase of the preheating will be identical to the base case of $m^2\phi^2$ potential with $g^2\phi^2\chi^2$ interaction before the effects of the potential features alter the dynamics.
Consequently, as confirmed by linear fluctuation analysis, the $\chi$ sector initially drives the resonance in all cases.
This stronger resonance in $\chi$ is reflected in the initial growth of the ratio.
However, the production of $\delta\phi$ due to rescattering from $\chi$ particles via processes such as $\chi_k\chi_k \to \delta\phi_k\delta\phi_k$ lowers this ratio.
However, the rescattering of $\chi$ particles against the inflaton zero modes, $\chi_k\phi_0\to \chi_k\delta\phi_k$, also effectively produces both $\delta\phi_k$ and $\chi$ particles, which again increases the ratio.
Finally, this ratio starts to fall when the back reaction kicks in. 
The additional growth of $\delta\phi$ fluctuations due to self-resonance for the cases with $h \neq 0$ keeps the ratio close to one for a longer duration.
In fact, for a sufficiently large value of $h$, away from zero in both directions, the self-resonance effects will be very effective, and it will soon catch up with the resonance in the $\chi$-sector. 
This fact is seen from the faster reduction in the above ratio.
We can define the time when backreaction kicks as $\tilde{t}_{\mathrm{br}} \equiv mt_{\mathrm{br}}$ from the moment when $\chi_\mathrm{rms}$ reaches its maximum value.
For the five cases considered here, we find that $\tilde{t}_{\mathrm{br}} \approx (114,\,145,\,146,\,220,\,229)$ for $h=(0.815,\,0.4,\,0.0,\,-0.5,\,-0.815)$, respectively.
Comparing this with the fact that the effective curvature of the potential, as shown in Fig. \ref{fig:BG01}, decreases as $h$ is reduced, we see that this facilitates the inflaton losing energy more rapidly than the daughter field due to the presence of the bumps. 
Which, in turn, enhances the energy transfer to the decay products.

\subsection{Equation of the State parameter and distribution of energy}
\begin{figure}[!ht]
    \begin{center}
    \includegraphics[height=1.6cm]{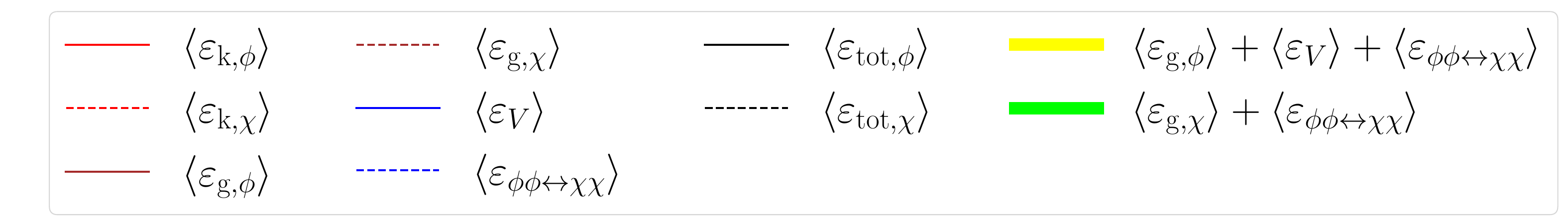}
    \begin{minipage}{0.48\textwidth}
        \centering
        \includegraphics[width=0.9\textwidth]{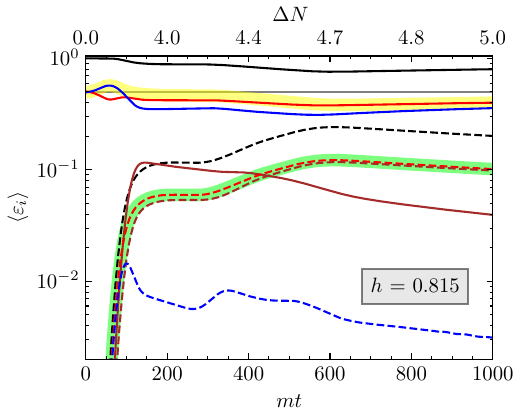}\\ 
        \includegraphics[width=0.9\textwidth]{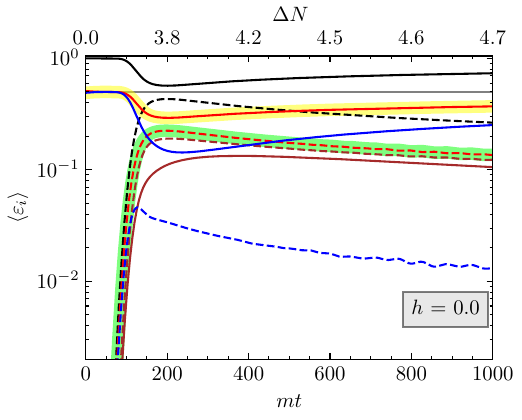}
    \end{minipage}
    \begin{minipage}{0.48\textwidth}  
        \centering
        \includegraphics[width=0.9\textwidth]{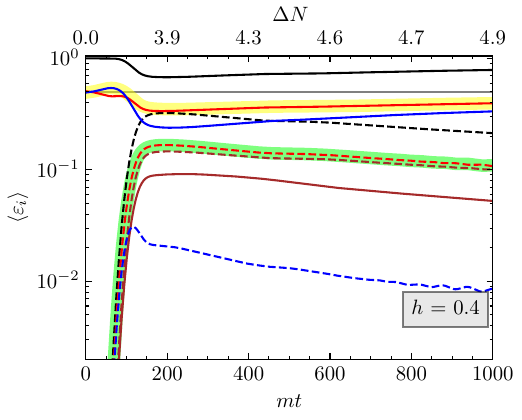}\\ 
        \includegraphics[width=0.9\textwidth]{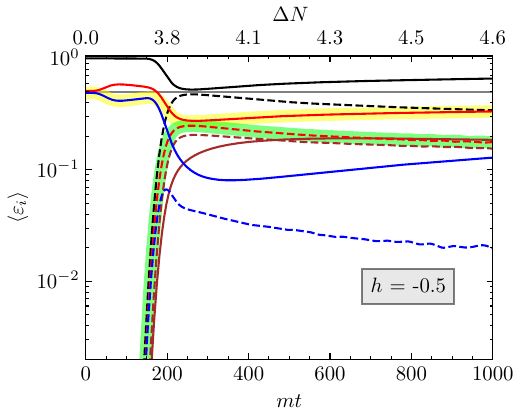} 
    \end{minipage}
    \includegraphics[width=0.45\textwidth]{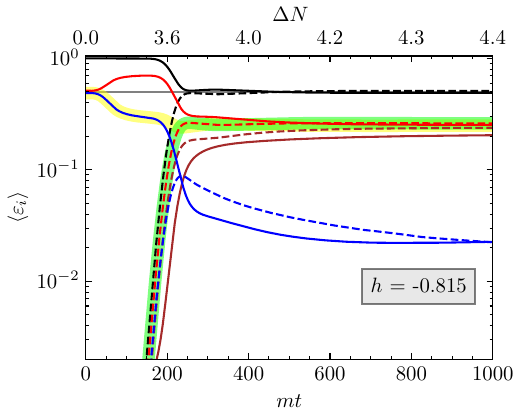}
    \end{center}
    \caption{
    We plot the lattice and oscillation average of different components of the energy densities normalized by the total energy density. 
    As we increase the height of the bumps, the system is essentially reduced to a single field dominated by the inflaton. 
    The dips increase the inflaton oscillation amplitude and help in more efficient interaction. 
    }
    \label{fig:figEnVir}
\end{figure}

The equation of state parameter is one of the most useful quantities in cosmic evolution. 
During the inflationary expansion, we must have $w<-1/3$. 
Macroscopically, it signals the initiation of the RD era after reheating when $w$ reaches a value of $1/3$. 
We define the instantaneous EOS for the scalar field system as:
\begin{equation}
    w = \frac{\sum_i p_{i}}{\sum_i \rho_{i}} = \frac{\rho_{\rm k, \phi} + \rho_{\rm k, \chi} -\frac{1}{3}(\rho_{\rm g, \phi} + \rho_{\rm g, \chi}) - (\rho_V + \rho_{\rm int})}{\rho_{\rm k, \phi} + \rho_{\rm k, \chi} + \rho_{\rm g, \phi} + \rho_{\rm g, \chi} + \rho_V + \rho_{\rm int}\phantom{abc}},
\end{equation}
where for a field $f=(\phi,\chi)$, different energy components are kinetic term $\rho_{{\rm k}, f} = \frac{1}{2}\dot{f}^2$, gradient term $\rho_{{\rm g}, f} = \frac{(\nabla f)^2}{2a^2}$, the potential term for the inflaton $\rho_V = V(\phi)$, and the interaction term $\rho_{{\rm int}, f}$. 
To understand intuitively how the dip/bump modifies the preheating dynamics and the EOS during this phase, let us summarize the characteristics of preheating for a monotonic power-law potential $V(\phi)\propto \phi^n$.
At the initial stage of coherent oscillation of the scalar condensate, the effective oscillation averaged EOS for the power-law potential is given by $w_{\mathrm{co}} = {(n-2)/(n+2)}$~\cite{Mukhanov:2005sc}. 
The general behavior after this coherent oscillating phase for such simple power-law potentials can be classified as:
\begin{enumerate}[(i)]
    \item $\bm{n=2}$: For quadratic models with four-legs interaction, the EOS initially increases from the coherent oscillating value of zero, reaching a maximum of around $w \sim 0.3$, before gradually returning to zero. 
    Such a fall in the EOS is quite expected since preheating in a simple $m^2\phi^2$ model with $g^2\phi^2\chi^2$ interaction describes (at the perturbative level) scattering between the $\phi$ and $\chi$ particles rather than the decay of $\phi$. 
    Therefore, although the massive $\phi$ component may remain under-abundant during the initial stages of preheating, it will eventually rise to dominate the energy density of the system.
    We may resort to three-legs interactions of the form $\sigma\phi\chi^2$ to facilitate the complete decay of inflaton. 
    In the case of trilinear interaction, the backreaction of the created particles also displaces the global minimum of the potential away from $\phi = \chi = 0$. 
    The value of this non-zero VEV $\langle\phi\rangle$ using the Hartree approximation is found to be $\langle\phi\rangle\simeq -(\sigma/2)\langle\chi^2\rangle/(m^2 + g^2\langle\chi^2\rangle)$ with $\langle\chi^2\rangle$ being the variance of $\chi$. 
    This negative average amplitude of the inflaton reduces the contribution of the massive inflaton component to the total potential due to the $\phi\chi^2$ term.
    Thus, overall, we achieve a plateau behavior in the EOS for trilinear interaction~\cite{Dufaux:2006ee}.
    Moreover, The quantum decay of the $\phi$ field ($\phi\to\chi\chi$) will eventually result in the complete depletion of its energy.
    Consequently, a radiation-dominated epoch will eventually commence after such perturbative decay processes. 
    However, the EOS does not transition to the radiation-dominated EOS when considering only nonlinear preheating effects studied through classical lattice simulations.
    Thus, a perturbative decay is generally required to complete the reheating process for quadratic potentials.
    \item $\bm{2 < n < 4}$: An interesting behavior is observed for models $2 < n < 4$. 
    Although the energy density of $\phi$ still redshifts slower than the radiation, the EOS still reaches $1/3$ in a sufficiently later time as the inflaton fluctuations can now grow due to the self-resonance.~\cite{Lozanov:2016hid,Lozanov:2017hjm,Saha:2020bis,Antusch:2020iyq}.
    If interactions with other fields are present, then for a sufficiently large value of the resonance parameter, the initial growth of fluctuations will be due to the resonance in the daughter field sectors. However, at later times~---~when the resonance is relegated to the narrow resonance regime~---~the self-resonance will drive the system to a state where $w\sim 1/3$. This effect is exclusively due to the nonlinear effects.
    \item $\bm{n\geq 4}$: For scalar field models with $n\geq 4$, $w \to 1/3$ irrespective of the couplings to other fields~\cite{Lozanov:2016hid,Lozanov:2017hjm,Maity:2018qhi,Saha:2020bis,Antusch:2020iyq}. 
    The $\phi$ field energy density now redshifts faster (or as fast as radiation when $n=4$) than radiation; hence, the system will eventually be radiation-dominated.
\end{enumerate}
\begin{figure}[!ht]
    \begin{center}
    \begin{minipage}{0.48\textwidth}
        \centering
        \includegraphics[width=0.9\textwidth]{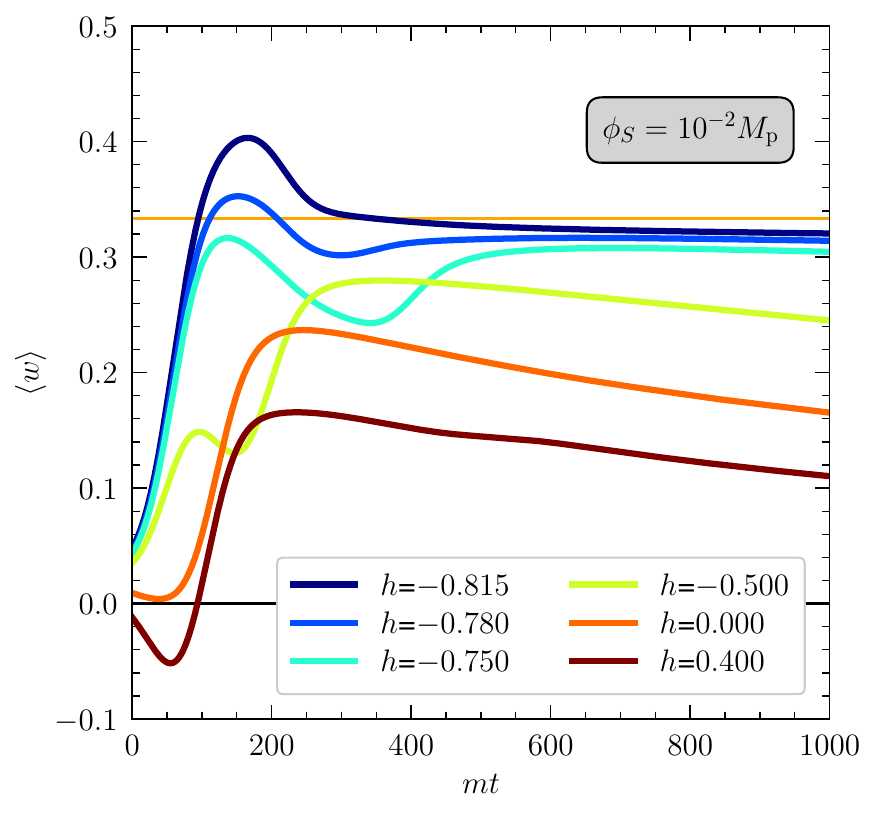} 
    \end{minipage}
    \begin{minipage}{0.48\textwidth}
        \centering
        \includegraphics[width=0.9\textwidth]{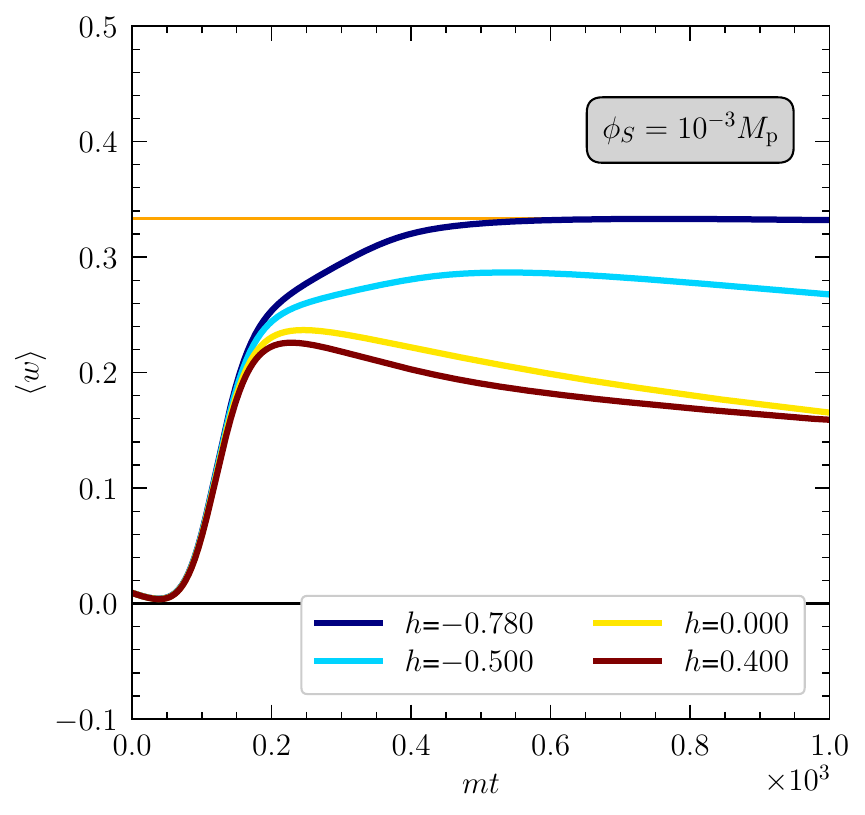} 
    \end{minipage}
    \end{center}
    \caption{
    We plot the evolution of the spatial averaged and oscillation averaged EOS $\langle w\rangle$ from the start of our simulation to some later time when the system reached a stationary phase.
    The EOS decreases after reaching the maximum value as preheating with four-legged interaction. 
    However, this decrease is gradual for the potentials with deformations and can reach a plateau-type behavior by moving the features closer to the minima. 
    The features are placed at $\phi_S =  10^{-2}\Mp$ while for the red dashed line, the features are placed at $\phi_S =  10^{-3}\Mp$.}
    \label{fig:figEoS}
\end{figure} 
The surge of higher power potential terms at small scales due to the features helps us gather the desired behavior in EOS without changing the potential during inflation.
Depending on the sign of the height of the feature, we can have a term adding or subtracting to the bare $m^2\phi^2$ potential. 
Contrary to the trilinear case, such terms are present at the background level, thus having more control over preheating dynamics. 
These features also modify the energy distribution among the components, as we describe below.
First, we plot the evolution of the EOS parameter for our model with features in the left panel of Fig~\ref{fig:figEoS} when the features are at $\phi_S=10^{-2}\Mp$. 
As we have described in Sec~\ref{sec:model}, increasing the height of the dips will widen the potential near the minima within a few $\sigma$'s around $\pm\phi_S$, which amounts to adding the higher-power contributions to the base term. 
These higher-power terms will make the EOS jump to their respective values at coherent oscillation. 
Now, depending on the value of $h$ and their positions, the effective surged higher-power terms can make it larger than $1/3$.
For the positions of the features as in the left panel of Fig \ref{fig:figEoS}, we see that as we start decreasing the height of the dip, the surged higher-power terms make the EOS rise from zero at a very early stage when the same for the base potential is still around zero.
This rise in the EOS is mostly due to the change of shape of the potential due to the surged higher-power terms, as the increase in the gradient term is still negligible at this stage (see the energy distribution among components below).
When the height is decreased further than $h\sim -0.75$, the EOS rises beyond $1/3$, which indicates the rise of potential terms larger than $n=4$.
However, as the Universe expands and the inflaton oscillation amplitude falls below the position of the features, the bare $m^2\phi^2$ term may again start dominating to bring down the EOS. 
Hence, eventually, the EOS reaches radiation-dominated EOS and keeps decreasing (at a rate determined by the shape of the potential via $h$) as the Universe expands. 
Now, by changing the position of the features, i.e., placing them closer to the potential minimum and restricting the height, we can achieve a situation when the EOS reaches the value of $1/3$ and keeps a constant value for a longer duration. 
The behavior of the EOS for this case is shown in the right panel of Fig~\ref{fig:figEoS}.
The reason for this stability is that as we move the features closer and closer to the minima, the effects of the features are appreciable for a much longer duration. At the same time, the growth of the gradient term makes the EOS stable against the further (very late) rise of the $m^2\phi^2$ term.
Further, to stabilize the EOS at late times, we can introduce trilinear terms~\cite{Dufaux:2006ee}, as we will comment in Appendix-\ref{app:trilinear}.
We have verified with very long-term simulations that for the chosen value of $\phi_S=10^{-3}\Mp$ and $h=-0.780$, the EOS parameter retains its value close to $1/3$.
In fact (as we will see in the next section), when the features are very close to the minima, the growth of inhomogeneity due to strong self-resonance at a later time is sufficient to drive the EOS to a radiation-like behavior.
Consequently, we will not focus on the trilinear terms in the main text.
These features also affect the distribution of the energy components and facilitate the energy transfer to the daughter field.
As expected in such systems, the system follows the virialization condition~\cite{Lozanov:2016hid,Lozanov:2017hjm,Figueroa:2016wxr,Maity:2018qhi}: $\langle\dot{f}^2\rangle = \langle|\nabla f|^2\rangle + \langle f(\partial V/\partial f)\rangle$, where the brackets denote the oscillation and spatial average. Writing the ratio of energy densities as 
\begin{align}
    \varepsilon_i \equiv \frac{\rho_i}{\sum_j \rho_j}\,,
\end{align}
the virial relation for each field can be written as:
\begin{align}
\label{eq:virial0}
    \langle\varepsilon_{\mathrm{k},\phi}\rangle &\simeq \langle\varepsilon_{\mathrm{g},\phi}\rangle + \frac{n}{2}\langle\varepsilon_{\phi}\rangle + \langle\varepsilon_{\mathrm{int}}\rangle,\\
    \langle\varepsilon_{\mathrm{k},\chi}\rangle &\simeq \langle\varepsilon_{\mathrm{g},\chi}\rangle + \langle\varepsilon_{\mathrm{int}}\rangle.
    \label{eq:virial1}
\end{align}
The evolution of different energy components is shown in Fig \ref{fig:figEnVir}.
 To better understand the status of the virial relations, we have taken the oscillation average of the volume-averaged energy components.  
 We can readily see the deviation from the simple quadratic case ($n=2$) in these figures during the evolution when the potential is dominated by the higher power terms.
 Earlier, we checked the emergence of higher-power terms from background $\phi$ evolution in Fig \ref{fig:BG01}.
 The deviation from the base $n=2$ case in the virial conditions, which now include the contribution from the total system, can be used to track the surge of the higher-power potential terms and when the massive component becomes dominant again.
 For this, we can rewrite the virial relation for $\phi$ in (\ref{eq:virial0}) as:
\begin{align}
n = \frac{\langle\varepsilon_{\mathrm{k},\phi}\rangle - \langle\varepsilon_{\mathrm{g},\phi}\rangle -\langle\varepsilon_{\mathrm{int}}\rangle}{\frac{1}{2}\langle\varepsilon_{\phi}\rangle}.
\label{eq:virialexpt}
\end{align}
\begin{figure}[!ht]
    \begin{center}
    \begin{minipage}{0.48\textwidth}
        \centering
        \includegraphics[width=0.9\textwidth]{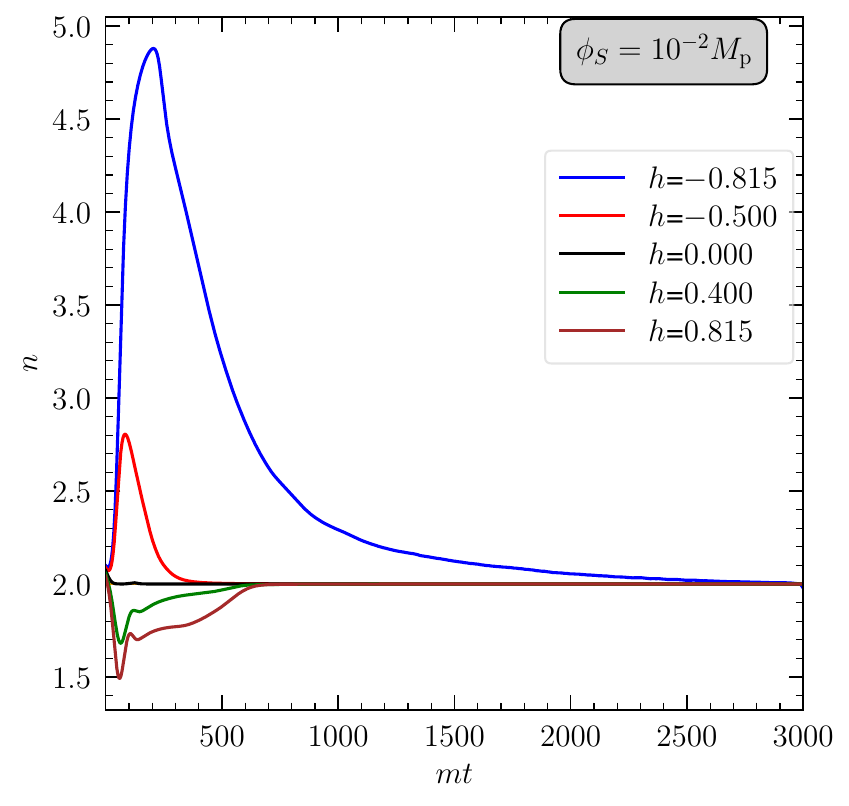} 
    \end{minipage}
    \begin{minipage}{0.48\textwidth}
        \centering
        \includegraphics[width=0.9\textwidth]{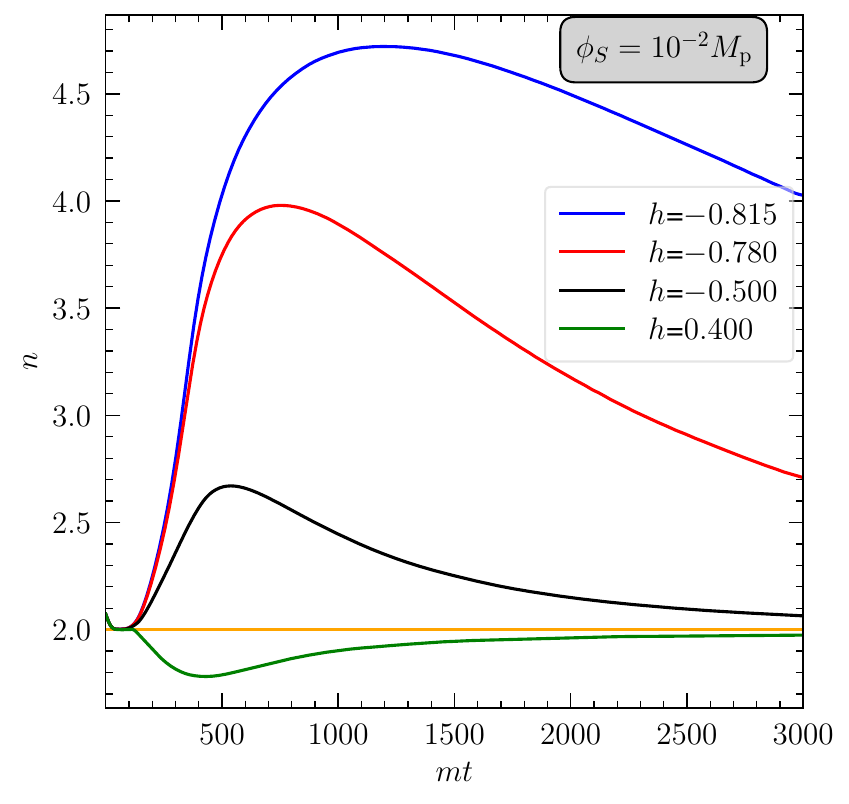} 
    \end{minipage}
    \end{center}
    \caption{
    We plot the evolution of the spatial and oscillation averaged combination of the energy densities in Eq.~(\ref{eq:virialexpt}) as a measure of the leading exponent of the potentials.
    }
    \label{fig:figVirExpt}
\end{figure}
We plot this combination in Fig~\ref{fig:figVirExpt} for two positions of the features.
When the features are at $\phi_S = 10^{-2}\Mp$, we see the dominance of terms greater than quadratic, confirming the earlier behavior from the background solutions as in Fig~\ref{fig:BG01}.
However, the quadratic term reemerges as the oscillation amplitude falls below the position of the features.
For a similar reason, the self-resonance in this case will also not be operative at later times.
Hence, we require resonance in the daughter fields to compensate for the consequent rapid fall of the equation.
Now, placing the features closer to the minima at $\phi_S = 10^{-3}\Mp$, the higher power terms will rise later but will dominate for longer. 
For the chosen set of parameters (so as not to change the global minima), we see that the leading power is around $n \lesssim 4$. 
 Therefore, in these cases (as we will see later), the self-resonance alone may be sufficient to drive the system to a radiation-dominated phase.
\par
It is also expected that the presence of the dip terms and the consequent surge of higher-power terms will help deplete the potential energy faster, converting it to kinetic energy that gets diluted faster. The energy transfer to kinetic terms and the growth of the gradient terms help in an equal distribution of energy among the two fields.
 We see this greater tendency to have an equipartition of energy among the two scalar components as we decrease the height beyond $h\sim 0.750$.
 Typically~---~as we have already mentioned, to prevent the re-emergence of the massive $\phi$-dominance~---~it is useful to have potential terms that will dilute energy faster than the massless $\chi$-component. 

\subsection{Energy transfer from inflaton to the daughter fields}
Preheating in a simple power-law potential with a four-legs interaction can make the energy distribute democratically among the different components~\cite{Figueroa:2016wxr,Maity:2018qhi,Antusch:2020iyq}.
The amount of energy transfer typically depends on the interaction between the components. 
For the quadratic model, for sufficiently large interaction, although a maximum of $50\%$ of the energy can be transferred to the $\chi$-component, the massive component will soon reemerge and dominate the system energy density. 
Naturally, simple power-law models with $n>2$ perform better in such cases, while on the other hand, the models with $n<2$ will be less effective in energy transfer to the daughter fields.
 \begin{figure}[!ht]
    \centering
\includegraphics[width=0.8\textwidth]{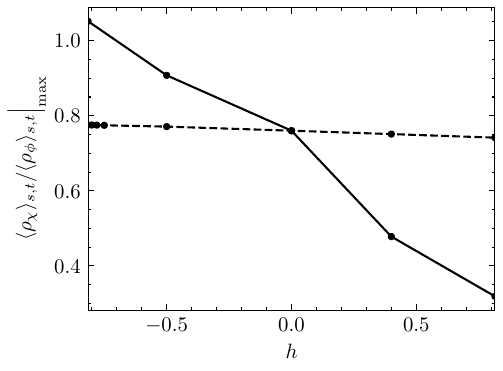}
    \caption{
    We plot the maximum value of the energy transferred from the inflaton sector to the $\chi$ sector as a function of the height $h$ of the features. The solid line is when $\phi_S=10^{-2}\Mp$ and the dashed line is for $\phi_S=10^{-3}\Mp$.
    When the features are very close to the minima, the two-field resonance is marginally improved, while the system is mainly driven by the self-resonance in the $\phi$-sector. 
    }
    \label{fig:figEnergyTransfer}
\end{figure}
These effects can also be seen for the potentials with dips/bumps, in which case the height $h$ acts as a knob to control the energy transfer. 
We plot the variation of the maximum energy transfer to the $\chi$ field with $h$ in Fig~\ref{fig:figEnergyTransfer} when the features are at $\phi_S = 10^{-2}\Mp$.
We see that the energy transfer almost monotonically increases as we decrease the value of $h$. 
In this work, as mentioned earlier, we will restrict the interaction strength to a value safe from any radiative correction to the inflationary potential at CMB scales.  
With such a typical value of interaction strength, we see that the maximum energy transfer is around $40\%$ for the base $m^2\phi^2$ case.
However, the enhanced resonance due to the emergence of the higher-power terms in the potential will lead to a greater transfer of the energy density to the $\chi$ component. Moreover, the re-emergence of the massive $\phi$ can be significantly delayed due to the rise in gradient term and enhanced resonance in the $\delta\phi$ sector.
For comparison, we have also plotted the energy transfer when the features are at $\phi_S = 10^{-3}\Mp$, which only marginally improves the energy transfer. This fact can be understood as when the features are very close to the minima; the self-resonance becomes more important while the two-field resonance essentially remains the same. 
In fact, in this case, the self-resonance may be sufficient to drive the resonance without the additional coupling to $\chi$.
In the next section, we will discuss the self-resonance when the system only has an inflaton component.

 \subsection{\label{sec:self_res}Lattice result for self-resonance with features}
\begin{figure}
    \centering
    \includegraphics[width=0.6\textwidth]{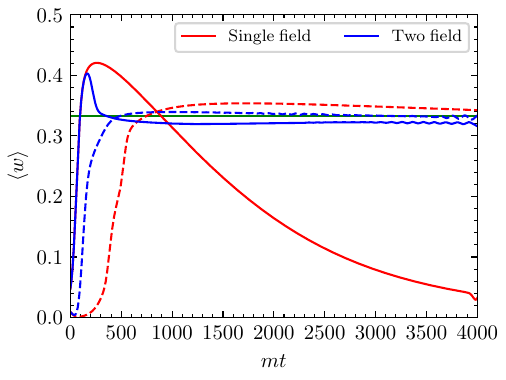}
    \caption{
    We plot the long-time behavior of the EOS to illustrate the effect of the position of the features in two-field resonance (blue) and single-field self-resonance (red). 
    We have set $h=-0.815$ for both cases and placed the features at $\phi_S = 10^{-2}\Mp$ (solid lines) and $\phi_S = 10^{-3}\Mp$ (dashed lines).
    For $\phi_S = 10^{-2}\Mp$, the EOS, during the coherent oscillation regime, jumps to a value close to $w=0.42$ due to higher-power potential terms. 
    However, without resonance in the $\chi$ field, the EOS gradually falls back to the matter-like EOS due to the re-emergence of the $m^2\phi^2$ term.
    For the two-field case, the growth of the gradient terms stabilizes the EOS at a value slightly less than $1/3$. 
    This later value of $w$ depends on the value of $h$ and can be slightly changed by decreasing $h$ further up to the point that does not introduce a new minimum other than at $\phi=0$.
    Now, when the features are closer to the minima~---~at $\phi_S = 10^{-3}\Mp$~---~the self-resonance is more important, as we have commented in the last section.
    The self-resonance and, consequently, the growth of the gradient term of the $\phi$ is sufficient to stabilize the EOS for a longer duration.}
    \label{fig:figEosStepStepsFlds}
\end{figure}

 We will now consider the results for the model in the absence of any interactions.
 As evident from the linear analysis in Sec~\ref{sec:pre_lin} and the lattice results from previous sections, the potentials will experience self-resonance when $h\neq 0$. In particular, as we have seen from Fig~\ref{fig:figVirExpt} and the subsequent discussion, the higher power terms in the potential due to the features will dominate the quadratic term for an extended duration when the features are relatively closer to the minimum.
 When the features are at $\phi_S=10^{-2}\Mp$, the self-resonance due to the surged nonlinear terms kicks the EOS to a value larger than $1/3$ early in the time evolution. 
 The initial growth of the gradient term also indicates the effect of self-resonance.
 However, as the inflaton moves away from the features, the effect of nonlinear terms diminishes as the potential returns to the quadratic behavior.
 Now, comparing this to the two-field case, we see the resonance in the daughter field drives the growth of inhomogeneities of the system. In contrast, the EOS monotonically returns to the matter-like behavior for the single-field case.
 Now, placing the features closer at $\phi_S=10^{-3}\Mp$, as we saw earlier, the dominance of the surged nonlinear terms will make the self-resonance more robust and delay the resurgence of the massive component.
 The gradient energy will contribute a significant fraction of the total energy density. These will stabilize the EOS to a value $w\sim 1/3$ for an extended period, as we have checked with a very long-term simulation.
 \begin{figure}[ht!]
\centering 
\includegraphics[height=1.5cm]{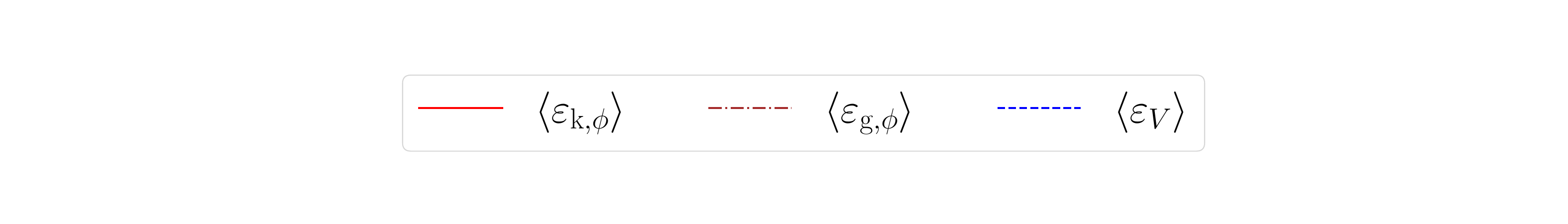}\par
\includegraphics[width=0.44\textwidth]{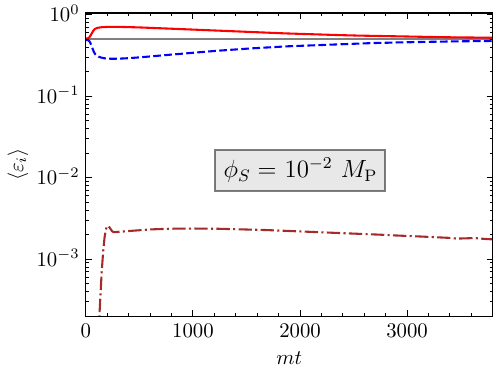}\quad
\includegraphics[width=0.44\textwidth]{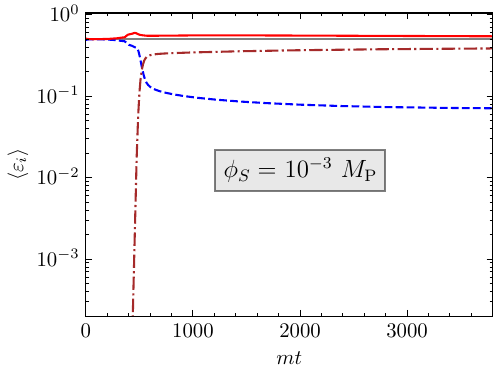}
\caption{
We show the evolution of different energy components for single-field self-resonance with two different positions of the features.}
\label{fig:figEnVirSingle}
\end{figure}
 The evolution of various energy components for the single field case with features as in the two field cases earlier are shown in Fig~\ref{fig:figEnVirSingle} for the height $h=-0.815$ while the behavior of the EOS compared to the two-field cases is illustrated in Fig~\ref{fig:figEosStepStepsFlds}.
 This phenomenon of self-resonance, even when the inflationary potential is given by a potential of the type $V(\phi)\propto\phi^2$, is intriguing. Such potentials may be useful for producing oscillons or similar exotic objects with minimal model building. 
 Moreover, this can enable us to study the non-perturbative particle production irrespective of the potential shape at the inflationary scales or even around the end of inflation.
 We will report some interesting results in these directions in future work.

\section{\label{sec:sec_gws}Generation of Gravitational Waves}
The exponential expansion during the Inflationary Universe and various non-perturbative processes in the post-inflationary epochs are known to generate stochastic gravitational wave background (SGWB) with distinct signatures.
These primordial GWs provide one of the cleanest signatures of the early Universe. 
The inflationary GWs are the quantum vacuum fluctuation of the metric during inflationary expansion, which gets stretched into classical GWs, affecting the B-mode polarization of the CMB radiation. 
On the contrary, the GWs produced during preheating result from the scattering of the classical inhomogeneities produced due to inflaton fragmentation. Hence, this provides us with a complementary channel to test the physics of the early Universe in a different frequency range, probing much smaller scale physics of the inflaton potential. 
Unlike other forms of radiation, gravitational waves pass through matter with minimal interaction, preserving the information of their origin. 
Additionally, due to cosmic expansion, the GWs can imprint the information of the equation of the state of various cosmological epochs through which it passes.
To study the dynamics of these GWs, in addition to the equations for the scale factor and the scalar fields in~(\ref{eq:phi})-(\ref{eq:H}), we need to solve the following linearized equation of motion of the tensor perturbations $h_{ij}$,
\begin{equation}
    \ddot{h}_{ij} + 3H\dot{h}_{ij} - \frac{\nabla^2}{a^2} h_{ij} = \frac{2}{\Mp^2}\Pi_{ij}^{TT},
    \label{eq:hij_eq}
\end{equation}
The tensor perturbations being transverse-traceless satisfy the following: $h_{ii} = \partial_ih_{ij} = 0$ and $\Pi_{ij}^{TT}$ being the transverse-traceless (TT) part of the effective anisotropic stress-energy tensor $\Pi_{ij} = \{\partial_i\phi\partial_j\phi + \partial_i\chi\partial_j\chi\}/a^2$. 
These metric equations are structurally similar to the scalar field equations and are solved using the same integration routine as the fields. The simulations are performed in an expanding but rigid background that neglects any backreaction from metric perturbations onto the field evolution.
The TT part of the tensor can be easily extracted in the Fourier space employing a projection operator (it is non-local in configuration space) defined as $\Pi_{ij}^{TT}(\bm{\mathrm{k}}) = \mathcal{O}_{ij,lm}(\hat{\bm{\mathrm{k}}})\Pi_{lm}(\bm{\mathrm{k}})$ with the projector $\mathcal{O}$ defined as
\begin{align}
    \mathcal{O}_{ij,lm}(\hat{\bm{\mathrm{k}}})\equiv P_{il}(\hat{\bm{\mathrm{k}}})P_{jm}(\hat{\bm{\mathrm{k}}}) - \frac{1}{2}P_{ij}(\hat{\bm{\mathrm{k}}})P_{lm}(\hat{\bm{\mathrm{k}}}),\quad P_{ij} = \delta_{ij} - \hat{\bm{\mathrm{k}}}_i\hat{\bm{\mathrm{k}}}_j,\quad\hat{\bm{\mathrm{k}}}_i = \bm{\mathrm{k}}_i/k
\end{align}
Once we have the solutions for the tensor perturbation, 
the associated energy density of gravitational waves is given by the $00$ component of the stress-energy tensor~\cite{Misner:1973prb}

\begin{equation}
    \rho_{\rm GW}(t) = \frac{\Mp^2}{4}\langle\dot{h}_{ij}(\textbf{x},t)\dot{h}_{ij}(\textbf{x},t)\rangle_{\mathcal{V}},
\end{equation}
where $\langle\cdots\rangle_{\mathcal{V}}$ denotes a spatial average over the lattice volume $\mathcal{V}$.
The spectrum of the energy density of GWs (per logarithmic momentum interval) observable today is~\cite{Easther:2006gt,Easther:2006vd,Easther:2007vj,Dufaux:2007pt}:
\begin{align}
\nonumber
h_0^2\Omega_{\rm GW,0} &= \frac{h_0^2}{\rho_{\mathrm{crit}}}\frac{\dx\rho_{\rm GW}}{\dx\log k}\Bigg|_{t=t_0} = \frac{h_0^2}{\rho_{\mathrm{crit}}}\frac{\dx\rho_{\rm GW}}{\dx\log k}\Bigg|_{t=t_\mathrm{e}}\frac{a_\mathrm{e}^4\rho_{\mathrm{e}}}{a_0^4\rho_{\mathrm{crit},0}}\\
&= \Omega_{\mathrm{rad}, 0}h_0^2\Omega_{\mathrm{GW},\mathrm{e}}\left(\frac{a_\mathrm{e}}{a_{\ast}}\right)^{1-3\weff}\left(\frac{g_{\ast}}{g_0}\right)^{-1/3},
\end{align}
where the subscript `$\mathrm{e}$' refers to the quantities evaluated at the end of the simulation, `$\ast$' is for the end of reheating, and `$0$' is for quantities evaluated today. 
The critical density of the Universe is $\rho_{\mathrm{crit}} = 3\Mp^2H^2$. 
We will also denote the dimensionless Hubble parameter today with a subscript `$0$' to avoid confusion with the height of the bumps/dips.
The equation of state~(EOS) of the Universe from the end of the simulation to the end of reheating is denoted by $\weff$. 
Notice that there is a redshift factor for $\weff \neq 1/3$, which depends on the total expansion during this period and the details of the reheating process.
The frequency of the produced GWs also redshifts similarly.
Now, substituting the value of the current radiation relic density,  we get the frequency and amplitude redshifted to today's observable values as~\cite{Saha:2022jdt,Cosme:2022htl}
\begin{align}
    f_{\mathrm{GW},0} &\simeq 4\times 10^{10} \mathbb{N}_{\mathrm{e}\to \mathrm{RD}}^{1/4}\frac{k}{a_{\mathrm{e}}H_{\mathrm{e}}}\left(\frac{H_{\mathrm{e}}}{\Mp}\right)^{1/2}~{\mathrm{Hz}},\\
    h_0^2\Omega_{\mathrm{GW},0} &\simeq 1.6\times10^{-5}\mathbb{N}_{\mathrm{e}\to \mathrm{RD}}\Omega_{\mathrm{GW,e}},
\end{align}
where we have denoted the redshift factor as $\mathbb{N}_{\mathrm{e}\to \mathrm{RD}} = \left(\frac{a_{\mathrm{e}}}{a_{\mathrm{RD}}}\right)^{1-3\weff}$. 
We have plotted the GW spectra that will be observed today for the models with different values of $h$ in Fig~(\ref{fig:figGWs}). We have neglected the redshift factor $\mathbb{N}_{\mathrm{e}\to \mathrm{RD}}$ for simplicity. We will roughly estimate the effect of this factor in the next section. 
The observed frequency is naturally in the ultra-high frequency range. 
Such high-frequency GWs are not detectable with the conventional current or planned detectors such as LIGO, DECIGO, etc. 
However, there are growing efforts and ideas for building such extremely high-frequency GW detectors~\cite{Berlin:2021txa,Ito:2022rxn,Ito:2023fcr,Capdevilla:2024cby}. 

The spectra show multiple prominent peaks, typical for a system with strong self-resonance with $n\geq 2$ potentials~\cite{Cui:2023fbg}. It provides a way to discern the change in the potential shape due to the features from their GW spectra. 
Moreover, since the features can be employed in other cases of GWs from the fragmentation of coherent scalars, such features in the base potential can even be traced with detectable GWs in the $\mathrm{MHz}$ to $\mathrm{nHz}$ range~\cite{Kitajima:2018zco,Saha:2022jdt,Cui:2023fbg}.
To supplement our result, in the next section, we will describe a more direct way to probe the presence of the features with GWs from the constraints on the effective number of relativistic degrees of freedom.

\subsection{\label{sec:GwsNeff}Constrain on \texorpdfstring{$\Delta N_{\mathrm{eff}}$}{TEXT}}
\begin{figure}[!ht]
    \begin{center}
    \begin{minipage}{0.48\textwidth}
        \centering
        \includegraphics[width=0.94\textwidth]{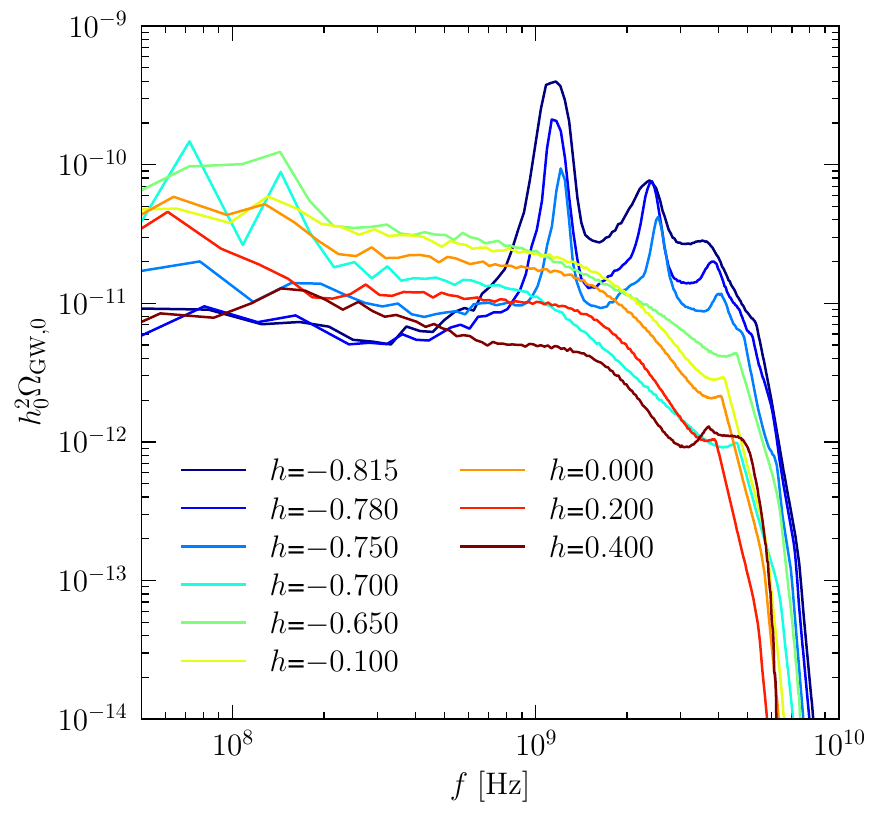} 
    \end{minipage}
    \begin{minipage}{0.48\textwidth}
        \centering
        \includegraphics[width=0.9\textwidth]{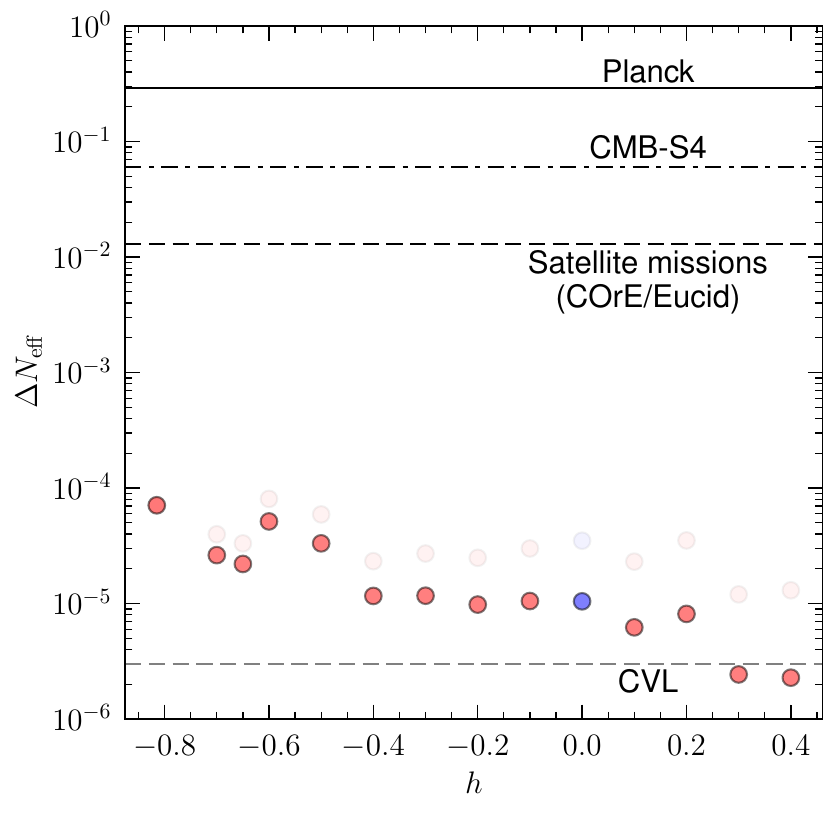} 
    \end{minipage}
    \end{center}
    \caption{
    In the left panel, we show the spectra of observed GWs for the base potential $h=0$ and for different values of $h$. The variation of the spectral shape with multiple prominent peaks that move towards higher frequency shows the surge of higher power potential terms.
    We plot the effective number of relativistic d.o.f from the Standard model as in (\ref{eq:neffGWs}) in the right panel, assuming the GWs only contribute towards its value. The transparent blobs are the actual value without considering the redshift factor when GWs are produced in a system with EOS other than $1/3$. The solid blobs are adjusted values assuming an expanding factor proportional to the EOS of the system at production, as described in the text. 
    }
    \label{fig:figGWs}
\end{figure}

The produced GWs in this phase are sub-hubble; consequently, their contribution to the energy density scales as radiation~\cite{Caprini:2018mtu}. 
It is thus possible to translate the GW contribution to the energy density of the universe into the constraints on the effective number of relativistic degrees of freedom. 
This number is typically parameterized by a deviation of the effective number of relativistic d.o.f from the Standard Model: $\Delta N_{\mathrm{eff}} =  N_{\mathrm{eff}} -  N_{\mathrm{eff, SM}}$,
with $ N_{\mathrm{eff},~SM}=3.0440$~\cite{Akita:2020szl,Froustey:2020mcq,Bennett:2020zkv}. If we assume that SGWB accounts for this additional d.o.f at the time of decoupling, we can write~\cite{Caprini:2018mtu}:
\begin{equation}
    \frac{h_0^2\Omega_{\mathrm{GW},0}}{h_0^2\Omega_{\gamma, 0}} = \frac{7}{8}\left(\frac{4}{11}\right)^{4/3}\Delta N_{\mathrm{eff}},
    \label{eq:neffGWs}
\end{equation} 
where the current value of photon energy density $h_0^2\Omega_{\gamma, 0} = 2.47\times10^{-5}$. 
The bound on $\Delta N_{\mathrm{eff}}$ from the Planck observations is $\left|\Delta N_{\mathrm{eff}}\right|\lesssim 0.29$ at $95\%$ C.L.~\cite{Planck:2018vyg,Pagano:2015hma}. 
Next-generation CMB-S4 experiments will be able to probe it at $\Delta N_{\mathrm{eff}}\lesssim 0.06$ at $2\sigma$~\cite{Abazajian:2019eic} while the satellite missions like COrE~\cite{COrE:2011bfs} and Eucild~\cite{EUCLID:2011zbd} will probe further $\Delta N_{\mathrm{eff}} \lesssim 0.013$ at $2\sigma$. 
We plot $\Delta N_{\mathrm{eff}}$ as a function of the deformation height in Fig~\ref{fig:figGWs}. 
The transparent points correspond to values without taking care of the redshift factor, while the actual points correspond to the values assuming that the scale factor has grown $20$ times from the end of the simulation to the time when the radiation domination is achieved. 
We can see from the figure that the $\Delta N_{\mathrm{eff}}$ varies almost linearly with the height of the potential feature. 
Although such contribution to $\Delta N_{\mathrm{eff}}$ solely from GWs contribution is unlikely to be detectable in the near future; however, the essential idea here is to showcase the effect of variation of the CMB observable due to small scale potential feature. 
With some futuristic proposals such as CVL~\cite{Ben-Dayan:2019gll}, it is possible to discriminate the small-scale features in the inflaton potential from $\Delta N_{\mathrm{eff}}$ observable. 

\section{\label{sec:disc_conc}Discussion and Conclusion}
In this work, we investigated how localized features in the inflationary potential influence the efficiency and dynamics of preheating. These small-scale features, giving rise to a surge of localized higher-power terms in the potential, can enhance resonance and energy transfer compared to the preheating in the base model. 
Importantly, these features do not affect the model predictions during the inflationary phase, preserving its consistency with CMB observations. 
By modeling these features as Gaussian terms symmetrically positioned around the potential minima at $\phi=0$, we find that the potential's curvature near the minima can be either reduced (for $h<0$, dip terms) or increased (for $h>0$, bump terms). It significantly modifies the preheating dynamics in a number of ways:
\begin{itemize}
    \item
    For $h<0$, the emergence of higher-power terms in the potential leads to resonance structures reminiscent of those observed in potentials with powers greater than two. 
    The slower reduction of the inflaton amplitude allows for a more prolonged period of interaction, facilitating efficient energy transfer to the daughter fields. 
    While self-resonance in the standard $m^2\phi^2$ potential happens only due to rescattering effects, the nonlinear terms in the potential naturally lead to self-resonance, and the overall effect is amplified.
    \item In contrast, for $h>0$, the narrower potential restricts the two-field resonance due to a faster decrease in the inflaton amplitude. 
    However, self-resonance becomes strong enough to drive preheating independently.
    \item The presence of the bumps in the potential also improves the energy transfer from the inflaton to the daughter fields. The initial dominance of higher-order terms in the potential leads to a more rapid conversion of potential energy into kinetic energy, which dilutes more quickly.
    Additionally, the enhanced growth of gradient terms contributes to a greater tendency toward equalizing the energy densities among the various components of the system.
    \item A particularly interesting observation is the dependence of the equation of state (EOS) parameter on the height and position of these features. With carefully chosen parameter values, the EOS can approach a radiation-dominated value during preheating and asymptotically stabilize at that level.
    \item These features leave detectable imprints on the generated GW spectrum sourced by the scalar fragmentation.
    Such imprints can be probed indirectly through $\neff$ observables and directly explored with the development of high-frequency GW detectors~\cite{Berlin:2021txa,Ito:2022rxn,Ito:2023fcr}. 
    Therefore, by combining the CMB-based observations for large-scale features with the GW detector results for small-scale features, it is possible to gain a comprehensive understanding of the complete shape of the inflationary potential.
\end{itemize}
\noindent
While we have focused on the implications of these features in the context of inflationary preheating, it is worth noting that they can have similar effects on the resonance and fragmentation of coherently oscillating scalars during any phase of the early Universe. For instance, such local modifications to the scalar potential can aid in studying the properties of any spectator scalars, such as axion-like particles, through the GW signals generated during the RD era~\cite{Kitajima:2018zco,Cui:2023fbg,Cui:2024vws}. 
These small-scale features, without altering any observables on the large scales, can profoundly affect the non-perturbative dynamics associated with parametric resonance. This opens a new avenue to investigate particle production from coherent, time-dependent classical scalar fields and related phenomena, including GW generation from scalar fragmentation.
\section*{Acknowledgements}
We thank Kyohei Mukaida for insightful comments. 
P.~S. and Y.~U. are supported by Grant-in-Aid for Scientific Research (B) under Contract No.~23K25873 (23H01177). 
Y.~U. is supported by Grant-in-Aid for Scientific Research under Contract Nos.~JP19H01894, JP21KK0050, and JST FOREST Program under Contract No.~JPMJFR222Y. 
Numerical computation in this work was carried out at the Yukawa Institute Computer Facility.

\appendix
\section{\label{app:tanh}A note on the base potential}
As we illustrated the main results using the archetypal $m^2\phi$-model, the features in the potential will qualitatively affect the preheating in a similar way for any inflationary model whose behavior around the minima is effectively given by a quadratic term. 
\begin{figure}[!h]
    \begin{center}
    \begin{minipage}{0.48\textwidth}
        \centering
        \includegraphics[width=0.9\textwidth]{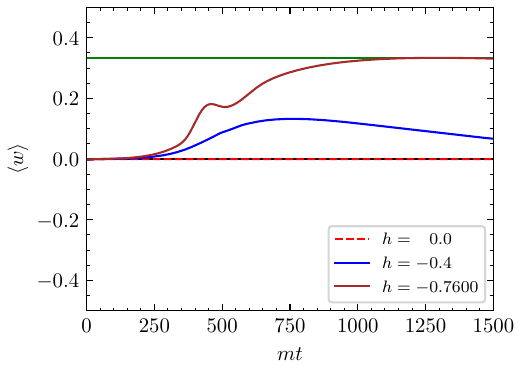} 
    \end{minipage}
    \begin{minipage}{0.48\textwidth}
        \centering
        \includegraphics[width=0.9\textwidth]{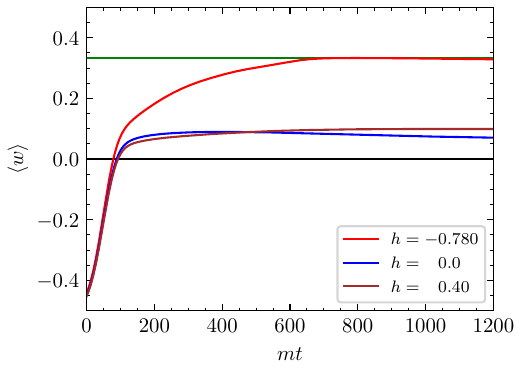} 
    \end{minipage}
    \end{center}
    \caption{
    We plot the evolution of the EOS for $\alpha$-attractor T-model in \ref{eq:PotTanh} for two values of the parameter $M$, which controls the width of the potential. The left panel is for $M=10\Mp$ when the preheating by self-resonance is inefficient for the base model. The figure on the right is for $M=7.75\times 10^{-3}\Mp$ when the self-resonance is somewhat efficient even for the base model. However, the EOS still does not reach the desired radiation-like value. Here, in the self-resonance case, the presence of features also provides the desired behavior.}
    \label{fig:figEoSTanh}
\end{figure} 
To complement our results, here we show the feature on one of the observational favored, the so-called $\alpha$-attractor T-model~\cite{Kallosh:2013hoa} with potential:
\begin{equation}
    \Lambda^4\tanh^{2n}\left(\frac{\phi}{M}\right).
    \label{eq:PotTanh}
\end{equation}
The scale $\Lambda$, which controls the height of the potential, is fixed by the amplitude of scalar spectra.
The scale $M$ determines the width of the potential, and it separates the quasi-flat part of the potential at large field values from a monomial term around the minima.
For large $M$~($M\approx \Mp$) and $n=1$, the potential can be identified with an effective quadratic potential; however, for smaller $M$ (reducing $M$ results in a smaller tensor-to-scalar ratio and hence corresponds to a more observationally favored region) the anharmonic terms are important for preheating dynamics.
Preheating in these models has been extensively studied in~\cite{Lozanov:2016hid,Lozanov:2017hjm,Antusch:2020iyq,Figueroa:2021iwm}.
We have shown the results in Fig (\ref{fig:figEoSTanh}) for two values of $M$ namely, (i) $M=10\Mp$, for which the 
potential around the minima behaves as $\mf^2\phi^2$. In this case, the self-resonance is not efficient. (ii) For sub-Planckian value of $M=7.75\times 10^{-3}\Mp$, the self-resonance is efficient.
The end of inflation in this model depends on the scale $M$, which determines the width of the potential.

\section{\label{app:trilinear}Effects of trilinear interaction}
As we have described earlier, at the perturbative level, the preheating with $\phi^2\chi^2$ four-legs interaction describes scattering between inflation quanta rather than the decay of inflation. 
Preheating produces copious amounts of inflaton quanta with non-zero momentum due to rescattering alongside inflaton at rest. 
This scattering between massive $\phi$ soon becomes inefficient, and to procure a complete decay of inflaton interactions of the form $\phi\chi^n$ are naturally considered. 
Although such complete decay would necessarily involve perturbative quantum processes, capturing such processes using classical lattice simulations may be difficult~\cite{Dufaux:2006ee}.
However, trilinear interactions are known to partly elevate the issues of preheating by helping to reach the equation of state a plateau behavior, as we have already mentioned in the introduction. The trilinear interactions can be more efficient in conjunction with the surge of higher-order potential terms with features,
Now, if inflaton is protected by a $Z_2$ symmetry, such interactions are forbidden. Despite, trilinear interactions may appear as a result of a spontaneous symmetry breaking $\phi\to\phi + \sigma$, for instance, in an inflation model with effective potential $V = -\frac{1}{2}m^2\phi^2 + \frac{1}{4}\lambda\phi^4 +  \frac{1}{2}g^2\phi^2\chi^2$ where the classical scalar VEV is $\sigma = m/\sqrt{\lambda}$. Nevertheless, in such cases, the mass of $\chi$-particles will be comparable to the $\phi$-mass that will complicate the decay of inflaton. 
\begin{figure}[!ht]
    \centering
\includegraphics[scale=0.5]{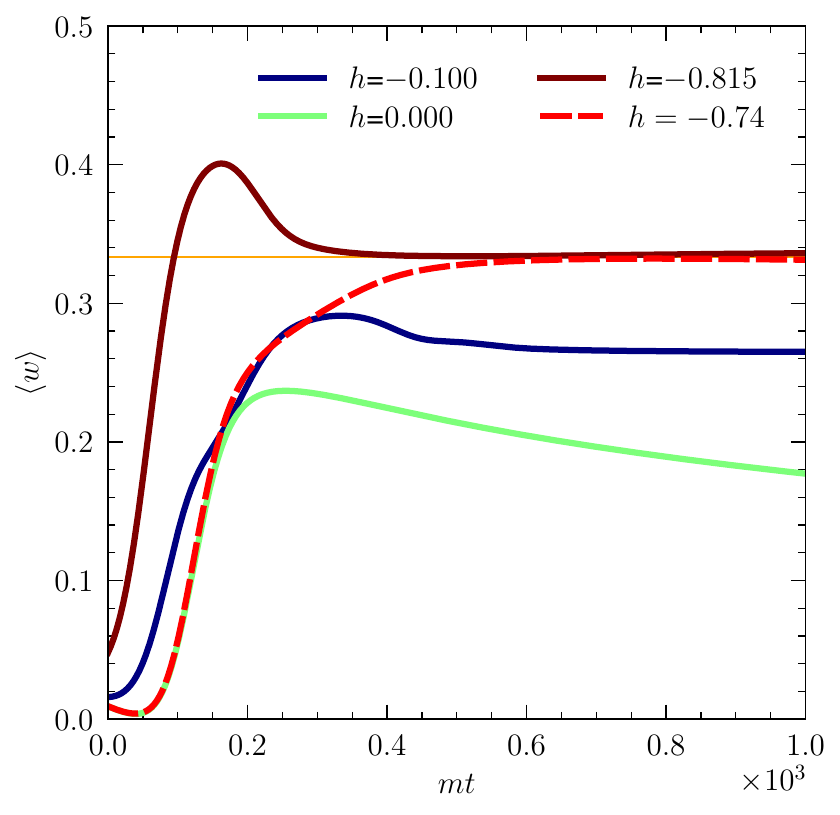}
    \caption{
    We plot the evolution of the EOS for different values of $h$. The trilinear terms, as expected, can stabilize the asymptotic behavior of the EOS.}
    \label{fig:figTrilinear}
\end{figure}
However, if we extend to the supersymmetric theories, we can naturally consider trilinear interaction~\cite{Dufaux:2006ee}.
In the subsequent discussion, following~\cite{Dufaux:2006ee}, we will consider trilinear interactions phenomenologically and extend their study to the case of potentials with features.
The scalar potential with trilinear interaction is given by: 
\begin{align}
    V(\phi,\chi) &= V(\phi) + \frac{1}{4}\lambda_{\chi}\chi^4 +  \frac{1}{2}g^2\phi^2\chi^2 +  \frac{1}{2}\sigma\phi\chi^2,
    \label{eq:trilinear}
\end{align}
where the $\chi^4$ self-interaction is now required to make the potential to be bounded from below. Working with the rescaled field and time variables as before, we identify the resonance parameters $q_4 = g^2\Phi^2_0/m^2$, $q_3 = \sigma\Phi_0/m^2$ and $q_{4,\chi} = \lambda_{\chi}\Phi_0^2/m^2$. We keep the potential bounded from below; we must have $\lambda_{\chi} > \sigma^2/2m^2$, which in terms of the resonance parameters reads $q_{4,\chi} > q_3/2$.
We plot the behavior of the EOS parameter for this model in Fig (\ref{fig:figTrilinear}) for $q_{4}^\mathrm{in} = 10^4$, $q_{3}^\mathrm{in} = 10$ and $q^{\mathrm{in}}_{4,\chi} = 50$ with $\phi_S= 10^{-2}\Mp$ for solid lines while the dashed line is for $\phi_S = 10^{-3}\Mp$ with other parameters remaining the same. 
As it was shown in~\cite{Dufaux:2006ee}, for the vanilla $m^2\phi^2$ with trilinear interaction to reach a plateau behavior for the EOS, we require a relatively larger value for the resonance parameters, namely $q^{\mathrm{in}}_4 = 10^4$, $q^{\mathrm{in}}_3 = 10^2$ and $q^{\mathrm{in}}_{4,\chi} = 10^4$ (for instance compare the behavior of the $h=0$ case). However, here, due to the presence of higher-order terms around the minima, the same level of efficiency can be reached with smaller resonance parameters. Moreover, contrary to the simple $m^2\phi^2$ with the trilinear case, we can reach radiation-like EOS, and the trilinear interaction provides the necessary stability for a later stage of evolution.

\section{Lattice simulation details}
Lattice simulation for preheating with the simple $m^2\phi^2$ model is perhaps one of the most studied cases among all scalar field models. 
This is not surprising as, apart from being the prototype model to study field theory, most of the models can be well approximated by a quadratic potential around the minima. 
However, since we have some non-trivial deviations from this simple behavior at small field values, it would be worth rechecking some convergence tests to validate the simulation and to make sure the numerical errors are in check. 
To this end, we first collect the equations that are being solved to track the energy conservation in the system. 
We refer the reader to the extensive documentation of $\bm{\mathcal{C}\texttt{osmo}\mathcal{L}\texttt{attice}}$ for further details~\cite{Figueroa:2020rrl,Figueroa:2021yhd}. 
The evolution of the scale factor is obtained by the following second-order equation (the Raychaudhuri equation):
\begin{equation}
    \frac{a''}{a} = \frac{a^2}{3}\left(\frac{\phiin}{\Mp}\right)^2\left[\tilde{\rho}_V + \tilde{\rho}_{\mathrm{int}} - 2\tilde{\rho}_{\mathrm{k},\phi} - 2\tilde{\rho}_{\mathrm{k},\chi}\right].
    \label{eq:raychoudhuri}
\end{equation}
To track the energy conservation of energy, the code uses the following equation (the Friedmann equation):
\begin{equation}
    a'^2 = \frac{a^2}{3}\left(\frac{\phiin}{\Mp}\right)^2\left[\tilde{\rho}_{\mathrm{k},\phi} + \tilde{\rho}_{\mathrm{k},\chi} + \tilde{\rho}_{\mathrm{g},\phi} + \tilde{\rho}_{\mathrm{g},\chi} +\tilde{\rho}_V + \tilde{\rho}_{\mathrm{int}}\right],
    \label{eq:frdm}
\end{equation}
where the \textit{tilde} over any variable is to denote the dimensionless \textit{program variables}, \textit{prime} denotes a derivative with respect to the \textit{program~time} as defined in (\ref{eq:rescaled_vars}) and the dimensionless \textit{program~densities} are defined as $\tilde{\rho}=\rho/(m^2\phiin^2)$.
\begin{figure}[!ht]
    \begin{center}
    \begin{minipage}{0.48\textwidth}
        \centering
        \includegraphics[width=0.9\textwidth]{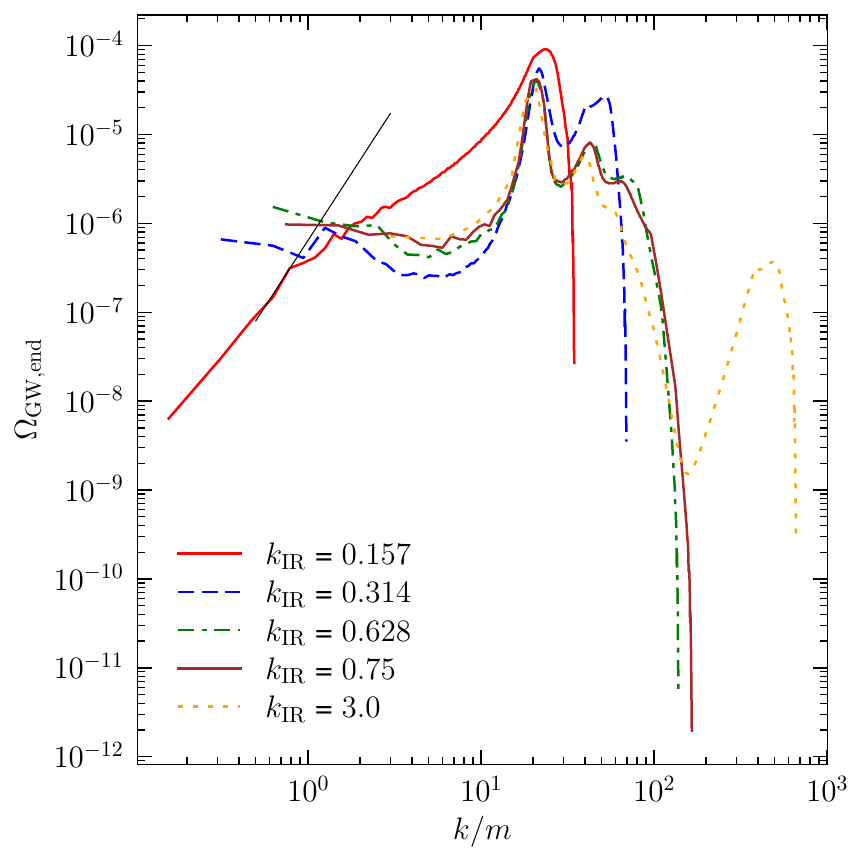} 
    \end{minipage}
    \begin{minipage}{0.48\textwidth}
        \centering
        \includegraphics[width=0.9\textwidth]{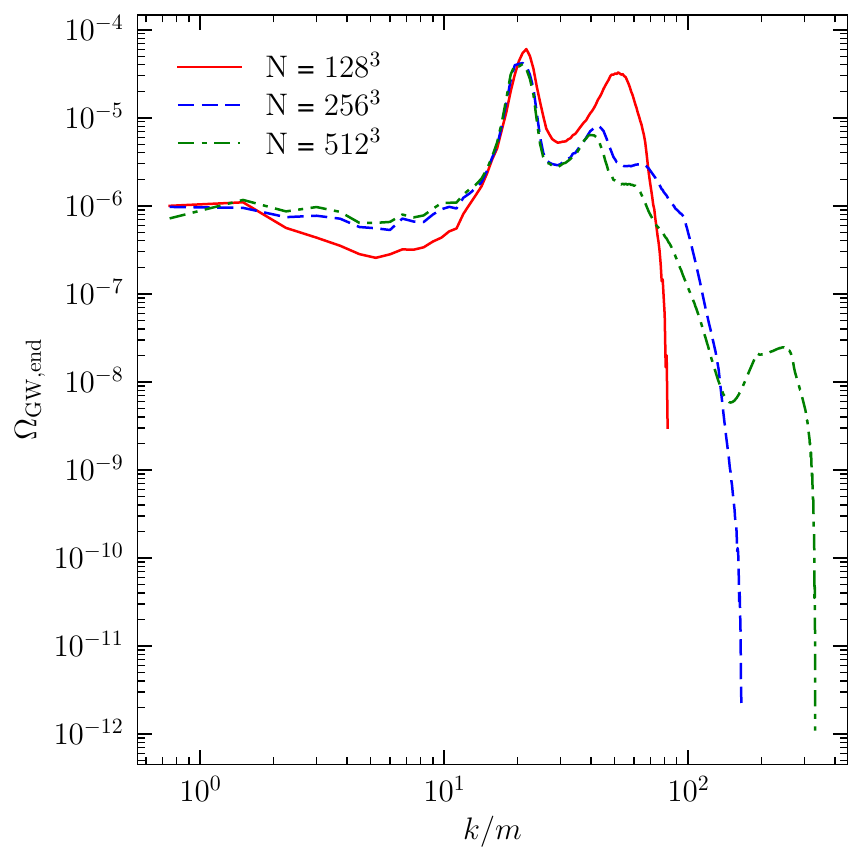} 
    \end{minipage}
    \end{center}
    \caption{
    We plot the convergence test for the GW spectrum at the end of the simulation. The thin black line is the $k^3$ line.
    }
    \label{fig:figConvergence}
\end{figure} 
Now, tracking the left-hand side (LHS) and right-hand side (RHS) of (\ref{eq:frdm}),
the code monitors the relative conservation of the Hubble constraints using the following output: $\langle~LHS-RHS\rangle/(\langle~LHS+RHS\rangle)$. In our simulation, we used the symplectic Leapfrog integration scheme and ensured that the above quantity remains smaller than at least the order of $10^{-5}$ during the simulation.
Additionally, since higher power terms add multiple peaks to the spectra, as expected from self-resonance in the inflaton sector, we have checked with different box sizes and number of grid points to ensure that the UV spectra are not plagued by numerical artifacts. 
The results of such convergence tests are shown in the figure below.
For the choice of model parameters and the initial values of the fields and momenta mentioned in the main text, the Hubble (in program unit) at the start of the simulation is found to be: $H_{\mathrm{pr}} = H/m \approx 0.5$; thus typically a (dimensionless) box with $L_{\mathrm{pr}}^{\mathrm{min}} = 1/H_{\mathrm{pr}} \sim 2$ or larger should be able to capture the physics we are concerned with. 
With this value of box size, the minimum (dimensionless) momentum mode that can be resolved is $k_{\mathrm{IR}} = 3.14$. So, ideally, we should choose a box containing several Hubble distances at the beginning of the simulation. 
First, we did a series of simulations with different values of $k_{\mathrm{IR}}$ with fixed lattice points ($N=256^3$). 
These results are shown in the right panel of Figure~\ref{fig:figConvergence}.
The smallest $k_{\mathrm{IR}}=0.157$ (corresponding to $L_{\mathrm{pr}}=40$) from these tests show that although it is able to capture the usual scaling behavior on large scales~\cite{Cai:2019cdl}, however, the UV part is prone to numerical artifacts. 
In this way, increasing the values $k_{\mathrm{IR}}$, we see that with $k_{\mathrm{IR}}=0.75$, we don't lose any important features in the spectra while also do not introduce any UV artifacts.
Next, with this value of $k_{\mathrm{IR}}=0.75$, we performed three simulations with different lattice points as shown in the right panel of Fig~\ref{fig:figConvergence} to find the optimal grid points for the simulations. 
In general, increasing the lattice points helps us regulate the UV part of the spectra. However, this also factors in with an increase in the run time of the simulation (for instance, doubling the lattice points will result in an $8$ times increase in the run time). 
Observing the spectra for three values of $N$, we took $N=256^3$ lattice points as a viable run parameter for the grid points. 


\providecommand{\href}[2]{#2}\begingroup\raggedright\endgroup

\end{document}